\documentclass[prd,aps,
nofootinbib,floatfix,
superscriptaddress]{revtex4}
\usepackage{graphicx}
\usepackage{epsfig}
\usepackage{rotating}
\usepackage{amssymb}
\usepackage{subfigure}
\usepackage{dsfont}
\usepackage{psfrag}
\usepackage{amsmath,euscript,array,mathrsfs}
\usepackage{slashed}
\usepackage{array}
\usepackage{youngtab}
\usepackage{color}
\usepackage{bbold}

\usepackage{hyperref}
\hypersetup{
colorlinks = true,
linkcolor = blue,
linktocpage = true,
citecolor = blue
}

\newcommand{\beq}{\begin{equation}}
\newcommand{\eeq}{\end{equation}}
\newcommand{\beqs}{\begin{eqnarray}}
\newcommand{\eeqs}{\end{eqnarray}}
\newcommand{\Tr}{{\rm Tr}}

\newcommand{\dd}{\mbox{d}}

\newcommand{\be}{\begin{equation}}
\newcommand{\ee}{\end{equation}}
\newcommand{\ba}{\begin{array}}
\newcommand{\ea}{\end{array}}

\newcommand{\BigCurlyLeft}{\left\{ \rule{-0.1cm}{0.75cm} \right.}
\newcommand{\BigCurlyRight}{\left. \rule{-0.1cm}{0.75cm} \right\}}

\begin{document}

\title{Toward minimal composite Higgs models from \\ regular geometries in bottom-up holography}
\author{Daniel Elander}
\affiliation{Montpellier, France}
\author{Ali Fatemiabhari}
\email{a.fatemiabhari.2127756@swansea.ac.uk}
\affiliation{Department of Physics, Faculty  of Science and Engineering,
Swansea University,
Singleton Park, SA2 8PP, Swansea, Wales, UK}
\author{Maurizio Piai}
\affiliation{Department of Physics, Faculty  of Science and Engineering,
Swansea University,
Singleton Park, SA2 8PP, Swansea, Wales, UK}

\date{\today}

\begin{abstract}

We study a bottom-up, holographic description of a field theory yielding the spontaneous breaking of an approximate $SO(5)$ global symmetry to its $SO(4)$ subgroup. The weakly-coupled, six-dimensional gravity dual has regular geometry. One of the dimensions is compactified on a circle that shrinks smoothly to zero size at a finite value of the holographic direction, hence introducing a physical scale in a way that mimics the effect of confinement in the dual four-dimensional field theory. We study the spectrum of small fluctuations of the bulk fields carrying $SO(5)$ quantum numbers, which can be interpreted as spin-0 and spin-1 bound states in the dual field theory. This work supplements an earlier publication, focused only on the $SO(5)$ singlet states. We explore the parameter space of the theory, paying particular attention to composite states that have the right quantum numbers to be  identified as pseudo-Nambu-Goldstone Bosons (PNGBs). 

We find that in this model the PNGBs are generally heavy, with masses of the same order as other bound states, indicating the presence of a sizeable amount of explicit symmetry breaking in the field theory side.
But we also find a qualitatively new, unexpected result. When the dimension of the field-theory operator inducing $SO(5)$ breaking is close to half of the space-time dimensionality, there exists a 
region of parameter space in which the PNGBs and the lightest scalar are both parametrically light in comparison to all other bound states of the field theory. Although this region is known to yield metastable classical backgrounds, this finding might be relevant to model building in the composite Higgs context.

\end{abstract}

\maketitle

\tableofcontents

\section{Introduction}
\label{Sec:introduction}

It has been just over ten years  since the discovery of the Higgs boson~\cite{Aad:2012tfa,Chatrchyan:2012xdj},
and in the intervening time  composite Higgs models (CHMs)~\cite{Kaplan:1983fs,Georgi:1984af,Dugan:1984hq},
in which
the Higgs fields originate as composite, pseudo-Nambu-Goldstone bosons (PNGBs) 
in a more fundamental theory, have
 gained  much attention in the literature---see
Refs.~\cite{Panico:2015jxa,Witzel:2019jbe,Cacciapaglia:2020kgq},
and the summary tables in Refs.~\cite{Ferretti:2013kya,Ferretti:2016upr,Cacciapaglia:2019bqz}.
Phenomenological and model-building
studies of realisations of this idea~\cite{
Katz:2005au,
Barbieri:2007bh,Lodone:2008yy,Gripaios:2009pe,Mrazek:2011iu,Marzocca:2012zn,
Grojean:2013qca,Barnard:2013zea,Cacciapaglia:2014uja,Ferretti:2014qta,Arbey:2015exa,
vonGersdorff:2015fta,Vecchi:2015fma,Cacciapaglia:2015eqa,Ma:2015gra,Feruglio:2016zvt,DeGrand:2016pgq,
Fichet:2016xvs,Galloway:2016fuo,Agugliaro:2016clv,Belyaev:2016ftv,Bizot:2016zyu,Csaki:2017cep,Chala:2017sjk,
Golterman:2017vdj,Csaki:2017jby,Alanne:2017rrs,Alanne:2017ymh,Sannino:2017utc,Alanne:2018wtp,Bizot:2018tds,
Cai:2018tet,Agugliaro:2018vsu,BuarqueFranzosi:2018eaj,Cacciapaglia:2018avr,
Gertov:2019yqo,
Ayyar:2019exp,
Cacciapaglia:2019ixa,
BuarqueFranzosi:2019eee,
Cacciapaglia:2019dsq,
Cacciapaglia:2020vyf,
Banerjee:2022izw,
Appelquist:2020bqj,
Cacciapaglia:2021uqh,
Appelquist:2022qgl,
Ferretti:2022mpy,Banerjee:2023ipb} have been complemented by a growing 
literature of dedicated lattice calculations, that analyse strongly-coupled field theories
providing, at least partial, short-distance completions~\cite{
Hietanen:2014xca,Detmold:2014kba,Arthur:2016dir,Arthur:2016ozw,
Pica:2016zst,Lee:2017uvl,Drach:2017btk,Drach:2020wux,Drach:2021uhl,
Bennett:2017kga,Bennett:2019jzz,Bennett:2019cxd,Bennett:2022yfa,
Ayyar:2017qdf,Ayyar:2018zuk,Ayyar:2018ppa,
 Ayyar:2018glg,Cossu:2019hse,Shamir:2021frg,DelDebbio:2021xlv}.
But providing a compelling  microscopic origin for 
CHMs with  minimal $SO(5)/SO(4)$ coset
is non-trivial---see for instance Ref.~\cite{Caracciolo:2012je}.

Within string theory and supergravity, 
it has been discovered that 
 gauge-gravity dualities, or holography~\cite{
Maldacena:1997re,Gubser:1998bc,Witten:1998qj,
Aharony:1999ti}, provide an alternative way to study special field theories in their non-perturbative regime.
Applications include
the holographic description of
 confinement~\cite{Witten:1998zw,Klebanov:2000hb,Maldacena:2000yy,Butti:2004pk},
the study of the  composite (glueball) mass spectra~\cite{Brower:2000rp,Wen:2004qh,
 Kuperstein:2004yf,Bianchi:2003ug,
Berg:2005pd,Berg:2006xy,Elander:2009bm,Elander:2010wd,Elander:2013jqa,
Elander:2014ola,Elander:2017cle,Elander:2017hyr},
chiral symmetry breaking ~\cite{Karch:2002sh,Babington:2003vm}, and masses of mesons~\cite{Sakai:2004cn,
Sakai:2005yt,Kruczenski:2003be,Nunez:2003cf,Erdmenger:2007cm}. 
But it is very challenging to embed within string theory and supergravity fully realistic
dynamical models yielding the low-energy theories relevant to CHMs. 
A few steps towards a top-down construction
for CHMs with $SO(5)/SO(4)$ coset have been taken recently~\cite{Elander:2021kxk}. A more general,  pragmatic,
bottom-up approach to holography exists,
in which the gravity dual is constructed classically, on the basis of ad hoc simplifying assumptions.
Indeed, much work on the minimal $SO(5)/SO(4)$ coset 
has been developed in this context, and make it a quite compelling scenario~\cite{Contino:2003ve,
Agashe:2004rs,
Agashe:2005dk,
Agashe:2006at,
Contino:2006qr,
Falkowski:2008fz,
Contino:2010rs,
Contino:2011np}.
Other CHMs suitable for lattice explorations have been the subject of recent
bottom-up holographic studies~\cite{Erdmenger:2020lvq,Erdmenger:2020flu,Elander:2020nyd,Elander:2021bmt}.

In this paper, we outline a new bottom-up holographic realisation of the
$SO(5)/SO(4)$ paradigm needed for minimal CHMs. 
The model is an extended  version of the very simple one 
that was studied in Ref.~\cite{Elander:2022ebt}, admitting  the same class of background geometries.
While Ref.~\cite{Elander:2022ebt} focuses on identifying regions of parameter
space relevant to understanding the physics of the dilaton, the approximate Goldstone boson associated
with scale invariance, along  the programmatic lines developed in Refs.~\cite{Kaplan:2009kr,Gorbenko:2018ncu,Gorbenko:2018dtm,
Elander:2020ial,Elander:2020fmv,Elander:2021wkc},\footnote{The common
 theme to this sequence of papers is that a light dilaton 
might emerge in the spectrum of strongly-coupled models if the dynamics brings them 
in proximity of (tachyonic) instabilities, the simplest holographic realisation of which 
is related to the Breitenlohner-Freedman (BF) unitarity bound~\cite{Breitenlohner:1982jf}.
But see also the critical discussion in Ref.~\cite{Pomarol:2019aae}, that proposes a bottom-up 
model in which the dilaton is not 
parameterically light, as well as the models of the earlier Refs.~\cite{Arean:2012mq,Arean:2013tja}, 
that do not yield a light dilaton at all.} 
in this publication and in future ones~\cite{vacuum,EWSB}
we are interested in investigating whether the PNGB states associated with the $SO(5)/SO(4)$
coset have model-building potential
in the CHM context.

Let us first  summarise the salient features of the model, borrowing results
from Ref.~\cite{Elander:2022ebt}. A  single scalar field is coupled to gravity in six dimensions,
and its dynamics is governed by a polynomial potential~\cite{Goldberger:1999uk,DeWolfe:1999cp,Goldberger:1999un,Csaki:2000zn,
ArkaniHamed:2000ds,Rattazzi:2000hs,Kofman:2004tk}.
A free parameter in the potential determines the dimensionality of the operator (deformation)
in the dual five-dimensional field-theory interpretation of the gravity model.
Furthermore, one of the space-like dimensions is a circle, its size shrinking along
the holographic direction, which introduces a smooth end of space in the regular geometry. As proposed in Ref.~\cite{Witten:1998zw}, 
in the dual field-theory interpretation 
this amounts to introducing a mass gap, in a way that mimics what happens for confining theories. Notice that this is the main difference with respect to Refs.~\cite{Contino:2003ve,
Agashe:2004rs,
Agashe:2005dk,
Agashe:2006at,
Contino:2006qr,
Falkowski:2008fz,
Contino:2010rs,
Contino:2011np}, namely the existence of a lift to a completely smooth geometry in six dimensions, which imposes constraints on the bulk profiles of the scalar fields of the five-dimensional gravity theory, obtained after reduction on the circle.

It has been shown in Ref.~\cite{Elander:2022ebt}
that a phase transition occurs in the theory. This result was obtained by applying holographic renormalisation~\cite{Bianchi:2001kw,
Skenderis:2002wp,Papadimitriou:2004ap} to compute the free energy.
Whether the regular backgrounds are stable depends on the magnitude
 of the field-theory deformation. As explained in the body of the paper, the size of such deformation, and its associated condensate, are both extracted from the profile of the bulk fields, but are not independent of one another, as they are constrained by the aforementioned regularity conditions.

The calculation of the spectrum of fluctuations of the bulk fields
 has a natural holographic interpretation in terms of bound states of the
dual theory. Ref.~\cite{Elander:2022ebt} reports such spectrum,
for states that carry no $SO(4)$ quantum numbers, computed by exploiting the powerful algorithmic process developed
in Refs.~\cite{Bianchi:2003ug,Berg:2005pd,Berg:2006xy,Elander:2009bm,Elander:2010wd,
Elander:2014ola,Elander:2017cle,Elander:2017hyr,Elander:2018aub,Elander:2020csd}.
Depending on the region of parameter space of interest, 
the lightest scalar fluctuation is found to be either a tachyon, or a generic, massive spin-0 state, or a parametrically light dilaton 
only in rather special cases.
In particular,  the light dilaton emerges only in regions of parameter space 
for which the regular backgrounds are metastable,
while, in the region of parameter space in which the regular geometry is stable, the lightest scalar 
may show some suppression of its mass, but this is quantitatively  a modest effect,
and never a parametric one. 
In the body of this paper, we provide some technical details that
are necessary to the exposition, in the interest of making the presentation self-contained, 
and of fixing the notation,  while referring  instead to Ref.~\cite{Elander:2022ebt}
 for extensive details and numerical results.

In this paper, we replace the singlet scalar 
with an $SO(5)$ vector multiplet, gauge the $SO(5)$ symmetry  in
the  six-dimensional gravity geometry compactified on a circle,
adopt the $R_{\xi}$ gauge as in Ref.~\cite{Elander:2018aub}, and compute
 the  mass spectrum of new states carrying $SO(4)$ quantum numbers.
In doing so, we make essential use of the fact that we identify the single scalar field
 of Ref.~\cite{Elander:2022ebt} with the absolute value of the 
$SO(5)$ multiplet field, in such a way that the latter obeys the same equations of motion as the former,
and hence we consider identical classical background solutions.
The approximate, global symmetry breaking pattern
$SO(5)\rightarrow SO(4)$ emerges in the dual field theory. 
Despite our interest in CHMs,
here we describe the theory in isolation, and we do not couple it to external, weakly-coupled, elementary  fields,
deferring the actual construction of CHMs to future publications~\cite{vacuum,EWSB}. While in the stable region of parameter space, none of the composite states can be made parametrically light, interestingly we find that there exists a metastable region, in which the spectrum contains parametrically light PNGBs, accompanied by a light pseudo-dilaton. This is suggestive, as it indicates the need to include a dilaton in the low-energy effective theory~\cite{
Migdal:1982jp,Coleman:1985rnk,Goldberger:2008zz}.
Even taken in isolation, the emergence of a dilaton has striking, potential phenomenological implications,
and is the subject of a vast literature---see for example Ref.~\cite{
Hong:2004td,Dietrich:2005jn,Hashimoto:2010nw,
Appelquist:2010gy,Vecchi:2010gj,Chacko:2012sy,Bellazzini:2012vz,Bellazzini:2013fga,Abe:2012eu,
Eichten:2012qb,Hernandez-Leon:2017kea,CruzRojas:2023jhw} and references therein.
If one extends the chiral Lagrangian to the dilaton effective field theory~\cite{
Matsuzaki:2013eva,Golterman:2016lsd,Kasai:2016ifi,Hansen:2016fri,Golterman:2016cdd,Appelquist:2017wcg,
Appelquist:2017vyy,Golterman:2018mfm,Cata:2019edh,Cata:2018wzl,Appelquist:2019lgk,Golterman:2020tdq,
Golterman:2020utm,Appelquist:2022mjb},
then
 the dilaton field might have an important role to play also
 in the construction of a  viable CHM---see for instance  Refs.~\cite{Appelquist:2020bqj,Appelquist:2022qgl}.

The paper is organised as follows.
We present the model in Sec.~\ref{Sec:model}, and describe the classical solutions of interest in 
Sec.~\ref{Sec:solutions}, borrowing relevant results from Ref.~\cite{Elander:2022ebt},
but dispensing with repeating technical details and intermediate results. We then compute the mass spectrum of
 the fluctuations of the system, focusing on the  states
 carrying $SO(4)$ quantum numbers. We 
compare the results to those for the singlets~\cite{Elander:2022ebt}, 
by exploring the three-dimensional parameter space.
We summarise the main results and outline future research directions in Sec.~\ref{Sec:outlook}.
We relegate to the Appendices technical details that are useful to reproduce our main original results.

\section{The model}
\label{Sec:model}

In this section we provide the weakly-coupled gravity description of the 
model we want to analyse, which is closely related to the one 
studied in Ref.~\cite{Elander:2022ebt}. The two-derivative bulk action describes gravity in $D=6$ dimensions, coupled to a real scalar field ${\cal X}$ transforming in the fundamental representation of a gauged $SO(5)$-symmetry.
We add  two
 boundaries in the radial direction, at  $\rho=\rho_1$ and $\rho=\rho_2$, respectively,
 and hence the action includes appropriate boundary-localised terms.
 The boundaries  have the only purpose of acting as regulators: physical results
 can be recovered by extrapolating to the limit in which the boundaries are removed.
 
 The bulk, gauged $SO(5)$ is broken to $SO(4)$ by a
 the non-trivial
 vacuum expectation value (VEV) of the combination
 $\phi\equiv \sqrt{\mathcal X^T\mathcal X}$---the field $\phi$ can be identified
with the one appearing in Ref.~\cite{Elander:2022ebt}.
 The (putative) dual four-dimensional field theory has a global $SO(5)$ symmetry, 
 inherited from the bulk $SO(5)$.
And its breaking  is generically interpreted as
an admixture of spontaneous and explicit breaking effects, due to the coupling and VEV of the operator dual to the bulk field $\phi$.
In the treatment of the bulk theory, we
adopt the $R_{\xi}$ gauge,
for which purpose we follow the procedure (and  notation)  in Ref.~\cite{Elander:2018aub},
which requires to add both bulk and
boundary terms, but we do not report them in this section.

\subsection{The six-dimensional action}
\label{Sec:6}

We first write the model in $D = 6$ dimensions. The field content consists of  gravity,  
 scalar fields $\mathcal X_\alpha$ transforming in the $5$ of the gauge group $SO(5)$, 
 and $SO(5)$ gauge fields $\mathcal A_{\hat M \, \alpha}{}^\beta$. 
 The six-dimensional space-time indexes are denoted by $\hat M = 0,1,2,3,5,6$, 
 while the components of the fundamental representation of $SO(5)$ are denoted by Greek indexes $\alpha = 1, \cdots, 5$. 
 The generators $t^A$ ($A=1,\cdots,10$) of $SO(5)$ are normalised so that
  $\Tr (t^A t^B) = \frac{1}{2} \delta^{AB}$. The action is
  \begin{align}
	\mathcal S_6 &= \mathcal S_6^{(bulk)} + \sum_{i=1,2} \mathcal S_{5,i} \,, \\
	\mathcal S_6^{(bulk)} &= \int \dd^6 x \sqrt{-\hat g_6} \, \bigg\{ \frac{\mathcal R_6}{4} - \frac{1}{2} \hat g^{\hat M \hat N}\left( D_{\hat M} \mathcal X \right)^T D_{\hat N} \mathcal X - \mathcal V_6(\mathcal X) - \frac{1}{2} \Tr \left[ \hat g^{\hat M \hat P} \hat g^{\hat N \hat Q} \mathcal F_{\hat M \hat N} \mathcal F_{\hat P \hat Q} \right] \bigg\} \,, \\
	\mathcal S_{5,i} &= (-)^i \int \dd^5 x \sqrt{-\tilde{\hat g}} \, \bigg\{ \frac{\mathcal K}{2} + \lambda_i(\mathcal X) + f_i\left( \tilde{\hat g}_{\hat M \hat N} \right) \bigg\} \bigg|_{\rho = \rho_i} \,,
\end{align}
where the bulk part is $\mathcal S_6^{(bulk)}$, and the two boundary actions
 $\mathcal S_{5,i}$, with $i=1,2$, are localised at the values $\rho = \rho_{1,2}$ of the  radial coordinate.
Our conventions are such that
 the six-dimensional metric $\hat g_{\hat M \hat N}$ has determinant $\hat g_6$, 
 and signature mostly $+$.  The six-dimensional Ricci scalar is $\mathcal R_6$.
 The induced metric on the boundaries is denoted as $\tilde{\hat g}_{\hat M \hat N}$, 
 the extrinsic curvature is $\mathcal K$, and it
 appears in the Gibbons-Hawking-York (GHY) term of the boundary actions.
  The terms denoted with  $f_i$ depend explicitly on the induced metric 
 on the boundary, as in Ref.~\cite{Elander:2022ebt}.

The covariant derivatives are defined as follows:
\beqs
	\left( D_{\hat M} \mathcal X \right)_\alpha &\equiv \partial_{\hat M} \mathcal X_\alpha + i g \mathcal A_{\hat M \, \alpha}{}^\beta \mathcal X_\beta \,, 
	\eeqs
and the field-strength tensors are
	\beqs
	\mathcal F_{\hat M \hat N \, \alpha}{}^\beta &\equiv 2 \left( \partial_{[\hat M} \mathcal A_{\hat N] \, \alpha}{}^\beta + i g \mathcal A_{[\hat M \, \alpha}{}^\gamma \mathcal A_{\hat N] \, \gamma}{}^\beta \right) \,,
\eeqs
where antisymmetrisation is defined as $[n_1 n_2] \equiv \frac{1}{2} \left( n_1 n_2 - n_2 n_1 \right)$.
The coupling $g$ is a free parameter.

Both the boundary potentials $\lambda_i(\mathcal X)$, as well as the bulk scalar potential $\mathcal V_6(\mathcal X)$, are taken to be $SO(5)$ invariant, and hence functions of the single variable $\phi\equiv \sqrt{\mathcal X^T\mathcal X}$. We adopt the explicit form of $\mathcal V_6(\phi)$ following Ref.~\cite{Elander:2022ebt}, by expressing it in terms of a superpotential ${\mathcal W}_6$ that satisfies the relation
\beq
\label{eq:VfromW}
	\mathcal V_6 = \frac{1}{2} \sum_{\alpha}\left(\frac{\partial {\mathcal W}_6}
	{\partial{\mathcal X}_{\alpha}}\right)^2	
	- \frac{5}{4} \mathcal W_6^2 \,,
\eeq
where the superpotential is given by
\beq
	\mathcal W_6 \equiv -2 - \frac{\Delta}{2} {\mathcal X}^T{\mathcal X} \,=\,-2 - \frac{\Delta}{2} \phi^2\,,
\eeq
and hence one finds that
\beq
	\mathcal V_6 = -5 - \frac{\Delta (5 - \Delta)}{2} \phi^2 - \frac{5 \Delta^2}{16} \phi^4 \,.
\eeq
We retain this elegant formulation only for convenience, even though neither is the model itself supersymmetric (there are no fermionic fields), nor do the backgrounds discussed in this paper originate from solving first-order equations derivable from the superpotential $\mathcal W_6$.

\subsection{Dimensional reduction}
\label{Sec:reduction}

One of the dimensions is a circle, parameterised by the angular variable $0 \leq \eta < 2\pi$. 
We adopt the (soliton) ansatz:
\begin{align}
	\dd s_6^2 &= e^{- 2\chi} \dd x_5^2 + e^{6\chi} \left(\dd \eta + \chi_M \dd x^M \right)^2 \,,
\end{align}
where the space-time index $M = 0,1,2,3,5$. 
The five-dimensional metric has the domain-wall form
\beq
	\dd s_5^2 = \dd r^2 + e^{2A(r)} \dd x_{1,3}^2 = e^{2\chi(\rho)} \dd \rho^2 + e^{2A(\rho)} \dd x_{1,3}^2 \,,
\eeq
and we dimensionally reduce the theory, so that the reduced action is then
\begin{align}
\label{eq:action5d}
	\mathcal S_5 &= \mathcal S_5^{(bulk)} + \sum_{i=1,2} \mathcal S_{4,i} \,, \\
	\mathcal S_5^{(bulk)} &= \int \dd^5 x \sqrt{-g_5} \, \BigCurlyLeft
	\frac{R}{4} - \frac{1}{2} g^{MN} \left[ 6 \partial_M \chi \partial_N \chi + \sum_{\alpha = 1}^5 \left( D_M \mathcal X \right)_\alpha (D_N \mathcal X)_\alpha + e^{-6\chi} \sum_{A = 1}^{10} \left( D_M \mathcal A_6 \right)^A (D_N \mathcal A_6)^A \right] \\
	& \hspace{1.2cm} - e^{-2\chi} \mathcal V_6
	 - \frac{1}{2} g^2 e^{-8\chi} \mathcal X^T \mathcal A_6^2 \mathcal X - \frac{1}{16} e^{8\chi} g^{MP} g^{NQ} F^{(\chi)}_{MN} F^{(\chi)}_{PQ} - \frac{1}{2} e^{2\chi} \Tr \left[ g^{MP} g^{NQ} \mathcal F_{MN} \mathcal F_{PQ} \right] \nonumber \\
	& \hspace{1.2cm} {
	- g^{MN} (i g) \chi_M \mathcal X^T \mathcal A_6 D_N \mathcal X
	- 2 e^{2\chi} g^{MN} g^{OP} \chi_M \Tr \left( \mathcal F_{NO} D_P \mathcal A_6 \right)
	}
	\nonumber \\
	& \hspace{1.2cm} {
	- \frac{1}{2} g^2 g^{MN} \chi_M \chi_N \mathcal X^T \mathcal A_6^2 \mathcal X
	+ e^{2\chi} g^{MP} g^{NQ} \chi_M \chi_N \Tr \left( D_P \mathcal A_6 D_Q \mathcal A_6 \right)
	}
	\nonumber \\
	& \hspace{1.2cm} {
	- e^{2\chi} g^{MN} g^{PQ} \chi_M \chi_N \Tr \left( D_P \mathcal A_6 D_Q \mathcal A_6 \right)
	}
	\BigCurlyRight \,, \nonumber \\
	\mathcal S_{4,i} &= (-)^i \int \dd^4 x \sqrt{-\tilde g} \, \bigg\{ \frac{K}{2} + e^{-\chi} \lambda_i(\mathcal X) + e^{-\chi} f_i(\chi) \bigg\} \bigg|_{\rho = \rho_i} \,.
\end{align}
The five-dimensional metric $g_{MN}$ has determinant $g_5$, the induced metric on
 the boundaries is $\tilde g_{MN}$, the five-dimensional Ricci scalar is $R$, and $K$ is the extrinsic curvature. 
The field strength for the vector $\chi_M$ is given by $F^{(\chi)}_{MN} \equiv \partial_M \chi_N - \partial_N \chi_M$.
We define $\mathcal A_6 \equiv \mathcal A_6^A t^A$, where $\mathcal A_6^A$ is a scalar that transforms
 in the adjoint representation of $SO(5)$, and originates from the sixth component of the gauge field in six dimensions.

We are interested in background solutions in which $\mathcal A_6 = 0$, $\mathcal A_M = 0$, $\chi_M = 0$, 
while the metric and the scalars $\mathcal X_\alpha$ and $\chi$ depend on the radial coordinate, $\rho$, only. 
The  background fields satisfy the equations of motion
\begin{align}
\label{Eq:bg1}
	\partial_\rho^2 \mathcal X_\alpha + (4 \partial_\rho A - \partial_\rho \chi) \partial_\rho \mathcal X_\alpha &=
	\frac{\partial{ \mathcal V_6}}{\partial {\mathcal X}_{\alpha}} \,, \\
	\label{Eq:bg2}
	\partial_\rho^2 \chi + (4 \partial_\rho A - \partial_\rho \chi) \partial_\rho \chi &= - \frac{\mathcal V_6}{3} \,, \\
	\label{Eq:bg3}
	3 (\partial_\rho A)^2 - \frac{1}{2} \partial_\rho \mathcal X_\alpha \partial_\rho \mathcal X_\alpha - 3 (\partial_\rho \chi)^2 &= - \mathcal V_6 \,,
\end{align}
with boundary conditions given by
\begin{align}
\label{eq:BCX}
	\left( \partial_\rho \mathcal X_\alpha - \frac{\partial \lambda_i}{\partial \mathcal X_\alpha} \right) \bigg|_{\rho_i} &= 0 \,, \qquad
	\left( 6 \partial_\rho \chi + \lambda_i + f_i - \frac{\partial f_i}{\partial \chi} \right) \bigg|_{\rho_i} = 0 \,, \qquad
	\left( \frac{3}{2} \partial_\rho A + \lambda_i + f_i \right) \bigg|_{\rho_i} = 0 \,.
\end{align}
For vanishing $f_i = 0$, one obtains solutions that lift to domain walls in $D = 6$ dimensions,
for which 
\beqs
{\cal A}&=&A-\chi\,=\,3\chi.
\eeqs

The solutions which we will be interested in break the $SO(5)$ symmetry to $SO(4)$, due to a non-trivial background profile of $\phi(\rho) \neq 0$. It is hence convenient to decompose  ${\mathcal X}_\alpha$ as  $5=1\oplus 4$, in terms of irreducible representations of $SO(4)$, which we denote as $\phi$ and $\pi^{\hat{A}}$, respectively. We use the parameterisation:
\beqs
\label{eq:Xdecomp}
	\mathcal X &= \exp \left[ 2 i \pi^{\hat A} t^{\hat A} \right] \phi \, \mathcal X_0 \,, \qquad \mathcal X_0 = (0, 0, 0, 0, 1)^T \,, 
\eeqs
and adopt the decomposition
\beqs
\label{eq:Adecomp}
	\mathcal A_{\hat M \, \alpha}{}^\beta &= \mathcal A_{\hat M}{}^{\bar A} \big( t^{\bar A} \big)_\alpha{}^\beta + \mathcal A_{\hat M}{}^{\hat A} \big( t^{\hat A} \big)_\alpha{}^\beta \,,
\eeqs
where $t^{\hat A}$ ($\hat A = 1,\cdots,4$) and $t^{\bar A}$ ($\bar A = 5,\cdots,10$) are, respectively, the 
broken and unbroken generators of $SO(5)$ with respect to $\mathcal X_0$. 
An example of such a basis of generators is given in Appendix~\ref{Sec:so5}.
The generators obey the normalisation conditions
$\Tr (t^{\bar A} t^{\hat B}) = 0$, $\Tr (t^{\hat A} t^{\hat B}) = \frac{1}{2} \delta^{\hat A \hat B}$, 
and $\Tr (t^{\bar A} t^{\bar B}) = \frac{1}{2} \delta^{\bar A \bar B}$.

As the boundary potentials $\lambda_i(\phi)$ are $SO(5)$ invariant, the boundary conditions for $\mathcal X_\alpha$ given in Eq.~\eqref{eq:BCX} become
\beq
	0  = \left( \Big[ \partial_\rho \phi - \frac{\partial \lambda_i}{\partial \phi} \Big] \frac{\mathcal X_\alpha}{\phi} + 2 i \partial_\rho \pi^{\hat A} (t^{\hat A})_\alpha{}^\beta \mathcal X_\beta \right) \bigg|_{\rho_i} \,,
\eeq
which are solved by imposing
\beq
	\partial_\rho \phi |_{\rho_i} = \frac{\partial \lambda_i}{\partial \phi} \Big|_{\rho_i} \,, \qquad \partial_\rho \pi^{\hat A} |_{\rho_i} = 0 \,.
\eeq
These boundary conditions select background solutions in which, without loss of generality, we choose
$\pi^{\hat A} = 0$. Hence, the only background functions that are non-zero are $A$, $\phi$ and $\chi$.

\subsection{Truncation to quadratic order}
\label{Sec:quadratic}
 
 As $\phi$, $A$, and $\chi$ are the only functions that are non-trivial in the background,
 it is convenient to simplify the reduced action further, by power expanding the other scalar and gauge fields, and
 truncating the expansion at the quadratic order. The resulting action admits the same classical solutions, 
 and still contains enough information to 
compute the linearised equations of motion for the small fluctuations of all the fields.

By treating the remaining degrees of freedom (other than $\phi$, $\chi$, and $g_{MN}$) as perturbations, at quadratic order the five-dimensional action 
can then be written as
\beq
	\mathcal S_5^{(2)} = \mathcal S_5^{(bulk,2)} + \sum_{i=1,2} \mathcal S_{4,i} \,,
\eeq
where the bulk action is
\begin{align}
\label{eq:sigmamodelaction}
	\mathcal S_5^{(bulk,2)} =& \int \dd^5 x \sqrt{-g_5} \, \bigg\{ \frac{R}{4} - \frac{1}{2} g^{MN} G_{ab} \partial_M \Phi^a \partial_N \Phi^b - \mathcal V_5(\Phi^a) \nonumber \\ & - \frac{1}{2} g^{MN} G_{ab}^{(0)} \partial_M \Phi^{(0)a} \partial_N \Phi^{(0)b} - \frac{1}{2} m_{ab}^{(0)2} \Phi^{(0)a} \Phi^{(0)b} \nonumber \\
	& -\frac{1}{2} g^{MN} G^{(1)}_{AB} {\cal H}^{(1)}_M{}^A {\cal H}^{(1)}_N{}^B -\frac{1}{4} g^{MO} g^{NP} H^{(1)}_{AB}F_{MN}{}^{A}F_{OP}{}^B \bigg\} \,,
\end{align}
and the boundary actions are
\beq
	\mathcal S_{4,i} = (-)^i \int \dd^4 x \sqrt{-\tilde g} \, \bigg\{ \frac{K}{2} + e^{-\chi} \lambda_i(\phi) + e^{-\chi} f_i(\chi) \bigg\} \bigg|_{\rho = \rho_i} \,.
\eeq
The sigma-model metric for the active scalars $\Phi^a = \{ \phi, \chi \}$ is $G_{ab} = {\rm diag}\left(1,6\right)$,
and the potential is $\mathcal V_5(\phi,\chi) = e^{-2\chi} \mathcal V_6(\phi)$. The scalars $\Phi^{(0)a} = \{ \mathcal A_6^{\bar A}, \mathcal A_6^{\hat A} \}$ have sigma-model metric
\beqs
G^{(0)} = \left(\begin{array}{c|c}
e^{-6\chi} \mathbb 1_{6 \times 6} &\cr
\hline
&e^{-6\chi} \mathbb 1_{4 \times 4} \cr
\end{array}\right) \,.
\eeqs
and mass matrix
\beqs
\frac{m^{(0)2}}{g^2} = \left(\begin{array}{c|c}
\mathbb 0_{6 \times 6} &\cr
\hline
&\frac{1}{4} \phi^2 e^{-8\chi} \mathbb 1_{4 \times 4} \cr
\end{array}\right) \,.
\eeqs
The 1-forms $V_M{}^A = \{ \chi_M, \mathcal A_M{}^{\bar A}, \mathcal A_M{}^{\hat A} \}$ have field strengths $F_{MN}{}^A = 2\partial_{[M}V_{N]}{}^{A}$, and 
\beqs
H^{(1)} = \left(\begin{array}{c|c|c}
\frac{1}{4}e^{8\chi}&&\cr
\hline
&e^{2\chi}\,\mathbb{1}_{6 \times 6}&\cr
\hline
&& e^{2\chi}\,\mathbb{1}_{4 \times 4} \cr
\end{array}\right) \,,
\eeqs
while the gauge-invariant combinations of derivatives of the pseudo-scalars
 and 1-forms, given by ${\cal H}^{(1)}_M{}^A = \left\{ 0,0, \partial_M\pi^{\hat{A}}+\frac{g}{2} \mathcal A_M{}^{\hat{A}} \right\}$, have
\beqs
G^{(1)} = \left(\begin{array}{c|c|c}
0&&\cr
\hline
&\mathbb 0_{6 \times 6} &\cr
\hline
&& \phi^2 \,\mathbb{1}_{4 \times 4} \cr
\end{array}\right) \,.
\eeqs

\section{Background solutions}
\label{Sec:solutions}

All calculations presented in this paper make use of
regular background solutions in which the size of the circle shrinks to zero size. 
We refer to such solutions as confining, with abuse of language, and borrow their characterisation from 
Ref.~\cite{Elander:2022ebt}, to which we refer the Reader for technical details, and
expanded discussions.
The space of solutions of interest depends on two parameters.
The parameter $\Delta$ is related to the dimension of the deforming parameter, or operator, in the
five-dimensional theory. An additional parameter $\phi_I$ controls the behaviour of the active
scalars in proximity of the end of space, and ultimately controls the size of the deformation.

The solutions of interest are not known in closed form, but only numerically, and can be obtained
starting from the (IR) expansion of the background functions~\cite{Elander:2022ebt}. 
Assuming the space ends at some value $\rho_o$ of the radial direction, we can write 
the regular solutions as a power-expansion in the 
small difference $(\rho - \rho_o)$:
\beqs
\label{Eq:IR1}
	\phi(\rho) &=&
\phi_I - \frac{1}{16} \Delta \phi_I \left( 20 + \Delta 
	 \left( 5 \phi_I^2 - 4 \right) \right) (\rho - \rho_o)^2 
	 + \mathcal O\left((\rho - \rho_o)^4\right) \,, \\
	\label{Eq:IR2}
	\chi(\rho) &=&
 \chi_I + \frac{1}{3} \log(\rho - \rho_o) + \frac{1}{288} 
 \left( -80 + 8 \left( \Delta - 5 \right) \Delta \phi_I^2 - 5 \Delta^2 \phi_I^4 \right) 
 (\rho - \rho_o)^2 + O\left((\rho - \rho_o)^4\right) \,, \\
	\label{Eq:IR3}
	A(\rho) &= &
A_I + \frac{1}{3} \log(\rho - \rho_o) + \frac{7}{576} \left( 80 +\Delta \phi_I^2 \left( 40 + \Delta
 \left( 5 \phi_I^2 - 8 \right) \right) \right) (\rho - \rho_o)^2 + \mathcal O\left((\rho - \rho_o)^4\right) \,,
\eeqs
where $\chi_I$, $A_I$ are additional integration constants, besides the aforementioned $\rho_o$ and $\phi_I$.
In order to avoid a conical singularity in the plane described by $\rho$ and $\eta$, we set $\chi_I=0$,
and it is shown explicitly in Ref.~\cite{Elander:2022ebt} that the curvature invariants up to quadratic order
 (in six dimensions) are finite for these solutions. 

For asymptotically large values of the radial coordinate $\rho$,  all the backgrounds of interest approach 
the geometry of AdS$_6$. One can hence expand the functional form of the background functions in powers of the small parameter 
$z\equiv e^{-\rho}$. 
The detailed form of such UV expansions depends non-trivially on the value of $\Delta$,
and an (incomplete) catalogue of examples can be found in the Appendix of Ref.~\cite{Elander:2022ebt},
which we do not reproduce here.
We denote the five integration constants as $\phi_J$, $\phi_V$, $\chi_U$, $\chi_5$, and $A_U$.
They appear in the background functions in the following general way:
\beqs
	\phi(z) &=& \phi_{J} z^{\Delta_J}\,+\,\cdots\, + \phi_{V} z^{\Delta_V} \,+\,\cdots \,, \\
	\chi(z) &=& \chi_U - \frac{1}{3} \log(z)  \,+\,\cdots + ({\chi_5}+\cdots) z^5  \,+\,\cdots \,, \\
	A(z) &=& A_U - \frac{4}{3} \log(z)  \,+\,\cdots \,.
\eeqs
In these expressions, $A_U$ and $\chi_U$ can be set to zero without loss of generality, 
by trivial redefinitions of the metric and coordinates in six dimensions, 
for example following the
procedure adopted in Ref.~\cite{Elander:2020ial}. As we are interested only in computing the 
spectrum of fluctuations in units of the mass of the lightest tensorial glueball, 
our results are not affected by these two parameters, and 
we will not discuss them any further.
In the expansions in Ref.~\cite{Elander:2022ebt}, $\chi_5$ is conventionally defined in such 
a way that if $\chi_5=0$, then $A=4\chi$.
Finally, the two parameters $\phi_J$ and $\phi_V$ appear in $\phi(z)$, as the coefficients of the
$z^{\Delta_J}$ and $z^{\Delta_V}$ terms of the expansion.\footnote{The limiting case $\Delta=5/2$
requires a generalisation. The expansion in this case is written explicitly in Ref.~\cite{Elander:2022ebt}.}
We adopt the convention that $\Delta_J=\min (\Delta,5-\Delta)$, and 
$\Delta_V=5-\Delta_J$, hence always interpreting $\Delta_V$ as the dimension of the operator
in the dual field theory corresponding to $\phi$, and $\Delta_J$ as the dimension of its coupling.
We refer the reader to Ref.~\cite{Elander:2022ebt} for more details, and for the calculation of the free energy
for a number of choices  of $\Delta$ and $\phi_I$.

\section{Mass spectrum of fluctuations}
\label{Sec:glueballs}

Ref.~\cite{Elander:2022ebt} reports the spectrum of fluctuations of
the  $SO(5)$ singlets, computed using the gauge-invariant formalism of Refs.~\cite{Bianchi:2003ug,Berg:2005pd,
Berg:2006xy,Elander:2009bm,Elander:2010wd,
Elander:2014ola,Elander:2017cle,Elander:2017hyr}, which allows to resolve the mixing between 
fluctuations of the  fields $\phi$, $\chi$,  and the
metric.
The resulting variables are denoted, respectively, as $\mathfrak{a}^1$, $\mathfrak{a}^2$, 
 and $\mathfrak{e}$.  
We denote as $\mathfrak{v}^1$ the fluctuations associated with $\chi_M$.
For the same background solutions, we now consider the  $SO(5)$ multiplets,
${\cal A}_6^{\bar{A}}$, ${\cal A}_{6}^{\hat{A}}$, ${\cal A}_M^{\bar{A}}$, ${\cal A}_{M}^{\hat{A}}$, and $\pi^{\hat{A}}$.
None of these additional fields develop VEVs, hence they do not mix with components of the background  metric.
Yet, because of  the presence of a bulk $SO(5)$ gauge symmetry, 
to compute the spectrum of  their fluctuations we elect to introduce 
the $R_{\xi}$ gauge, and to identify gauge-invariant physical combinations,
borrowing the formalism developed in Ref.~\cite{Elander:2018aub} (see also Ref.~\cite{Elander:2021kxk}).
The resulting gauge-invariant fluctuations are denoted as $\mathfrak{a}^3$, $\mathfrak{a}^4$, $\mathfrak{v}^2$,
$\mathfrak{v}^3$, and $\mathfrak{p}$, respectively.

We restrict the discussion to the $SO(5)$ multiplets,
and the   
equations they obey, rather than repeating details that can be found in Ref.~\cite{Elander:2022ebt}.
The equations of motion are the following:

\beqs
 0 &=& \bigg[ \partial_\rho^2 + \left( 4 \partial_\rho A - 7 \partial_\rho \chi \right) \partial_\rho - e^{2\chi - 2 A} q^2 \bigg] \mathfrak a^3 \,,\\
 0 &=& \bigg[ \partial_\rho^2 + \left( 4 \partial_\rho A - 7 \partial_\rho \chi \right) \partial_\rho - \frac{g^2 \phi^2}{4} - e^{2\chi - 2 A} q^2 \bigg] \mathfrak a^4 \,,\\
 0 &=& \bigg[ \partial_\rho^2 + (2 \partial_\rho A + \partial_\rho \chi) \partial_\rho - e^{2\chi - 2 A} q^2 \bigg] \mathfrak v^2 \,,\\
 0 &=& \bigg[ \partial_\rho^2 + (2 \partial_\rho A + \partial_\rho \chi) \partial_\rho - \frac{g^2 \phi^2}{4} - e^{2\chi - 2 A} q^2 \bigg] \mathfrak v^3 \,,\\
  0 &=& \bigg[ \partial_\rho^2 - \left( 2 \partial_\rho A + \partial_\rho \chi + \frac{2\partial_\rho \phi}{\phi} \right) \partial_\rho - \frac{g^2 \phi^2}{4} - e^{2\chi - 2 A} q^2 \bigg] \mathfrak p \,,
 \eeqs
 where $q^2\equiv \eta_{\mu\nu}q^{\mu}q^{\nu}$, and $q^\mu$ is the four-momentum.

We study numerically the solutions of these linearised fluctuations in the range $\rho_1\leq\rho\leq\rho_2$, with $\rho_1>\rho_o$.
 In principle, in order to recover the physical results,
  we should apply boundary conditions at $\rho=\rho_1$ and $\rho=\rho_2$, and then repeat the process
 by taking the $\rho_1\rightarrow \rho_o$ and $\rho_2\rightarrow +\infty$ limits, separately.
 To be more specific, for the scalars $\mathfrak{a}^{3}$ and $\mathfrak{a}^{4}$ we impose Dirichlet 
 boundary conditions at $\rho=\rho_1$ and $\rho=\rho_2$: $\mathfrak{a}^{3,4} |_{\rho_i} = 0$. 
 For the vectors $\mathfrak{v}^{2}$ and $\mathfrak{v}^{3}$ we impose Neumann boundary conditions at $\rho=\rho_1$ and $\rho=\rho_2$: $\partial_\rho \mathfrak{v}^{2,3} |_{\rho_i} = 0$.
 Conversely, the pseudoscalar  $\mathfrak{p}$ obeys Dirichlet
 boundary conditions
 for $\rho=\rho_1$ and Neumann for $\rho=\rho_2$~\cite{Elander:2020nyd,Elander:2021kxk}. 
  In practice, in order to improve the numerical convergence of this process,
 we make use of the asymptotic expansions, both in the IR and UV, of the general
 solutions of the fluctuation equations---see an example  in
 Appendix~\ref{sec:IRUVexpansions}. We impose upon them the aforementioned boundary conditions,
and require 
 continuity of the functions and their derivatives with the expansions thus constrained.
The system is over-constrained, yielding a discrete spectrum of values of $M^2 = -q^2$.
 We will discuss in a future publication how these conclusions are modified in the presence of 
  non-trivial boundary terms~\cite{vacuum}, which non-trivially parametrise the effect of coupling the theory
  to external fields.

\begin{figure}[th]
\begin{center}
\includegraphics[width=16cm]{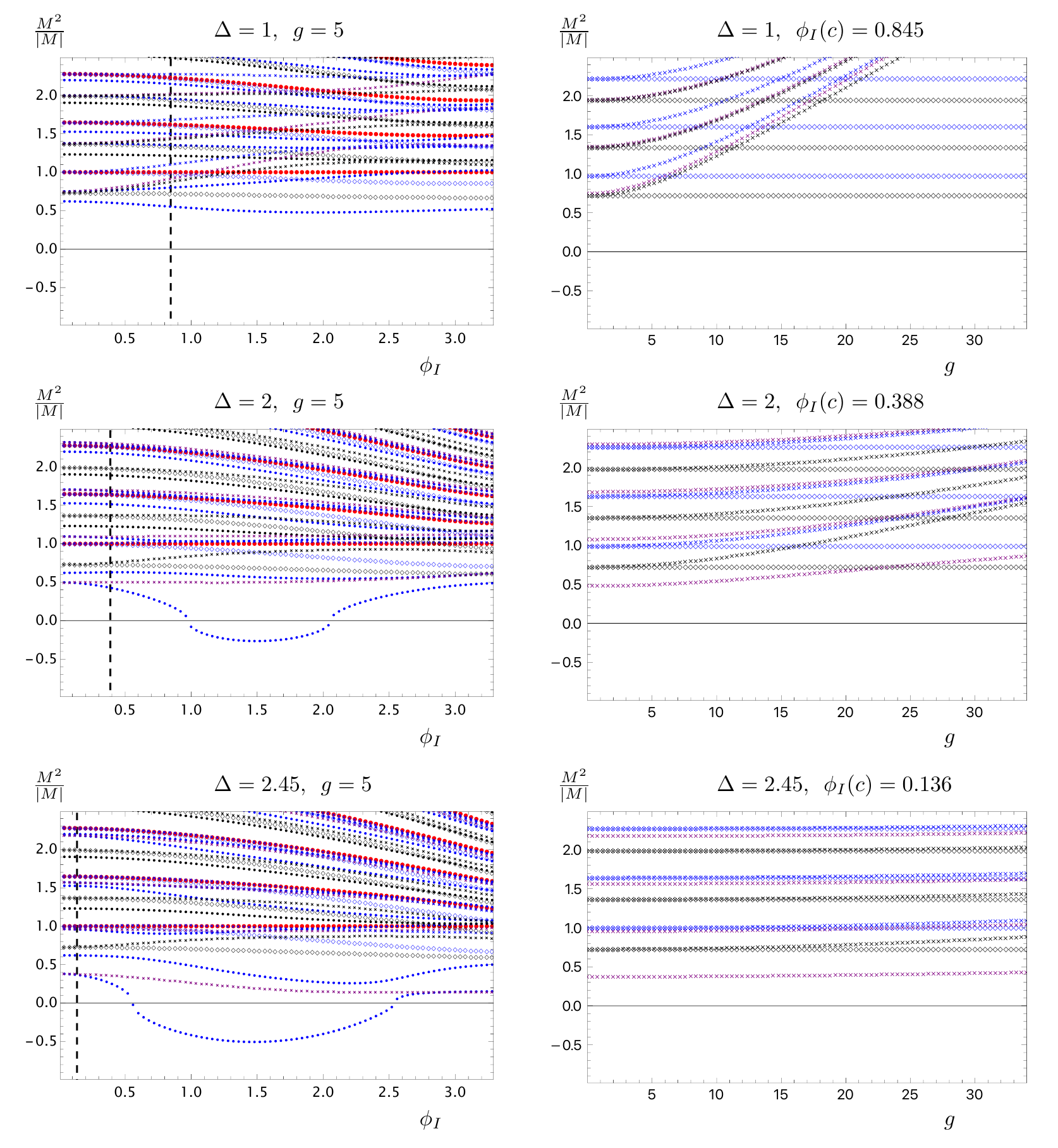}
\caption{Mass spectrum $\frac{M^2}{|M|}$ of fluctuations, computed for confining backgrounds, 
with various choices of $\Delta$, 
as a function of the IR parameter $\phi_I$ for $g=5$ (left),
and as a function of $g$ for $\phi_I=\phi_I(c)$ (right).
For each $\Delta$, we show the spectrum of scalar (blue), pseudo-scalar (purple),
vector (black), and tensor (red) states.  The values of the IR and UV cutoffs in the calculations are respectively given by $\rho_1 - \rho_o = 10^{-9}$ and  $\rho_2 - \rho_o = 5$, in all of the cases.
The different symbols refer to the quantum numbers in respect to the unbroken $SO(4)$ symmetry:
disks are used for singlets, and have already been reported in Ref.~\cite{Elander:2022ebt},
diamonds represent the $6$ of $SO(4)$, and crosses the $4$ of $SO(4)$.
All masses are normalised to the mass of the lightest spin-2 state.
Because the masses of the $SO(5)$ singlets do not depend on $g$, we do not repeat them in the right panels, 
which display only non-trivial $SO(5)$ multiplets.}
\label{Fig:mass1}
\end{center}
\end{figure}

\begin{figure}[th]
\begin{center}
\includegraphics[width=16cm]{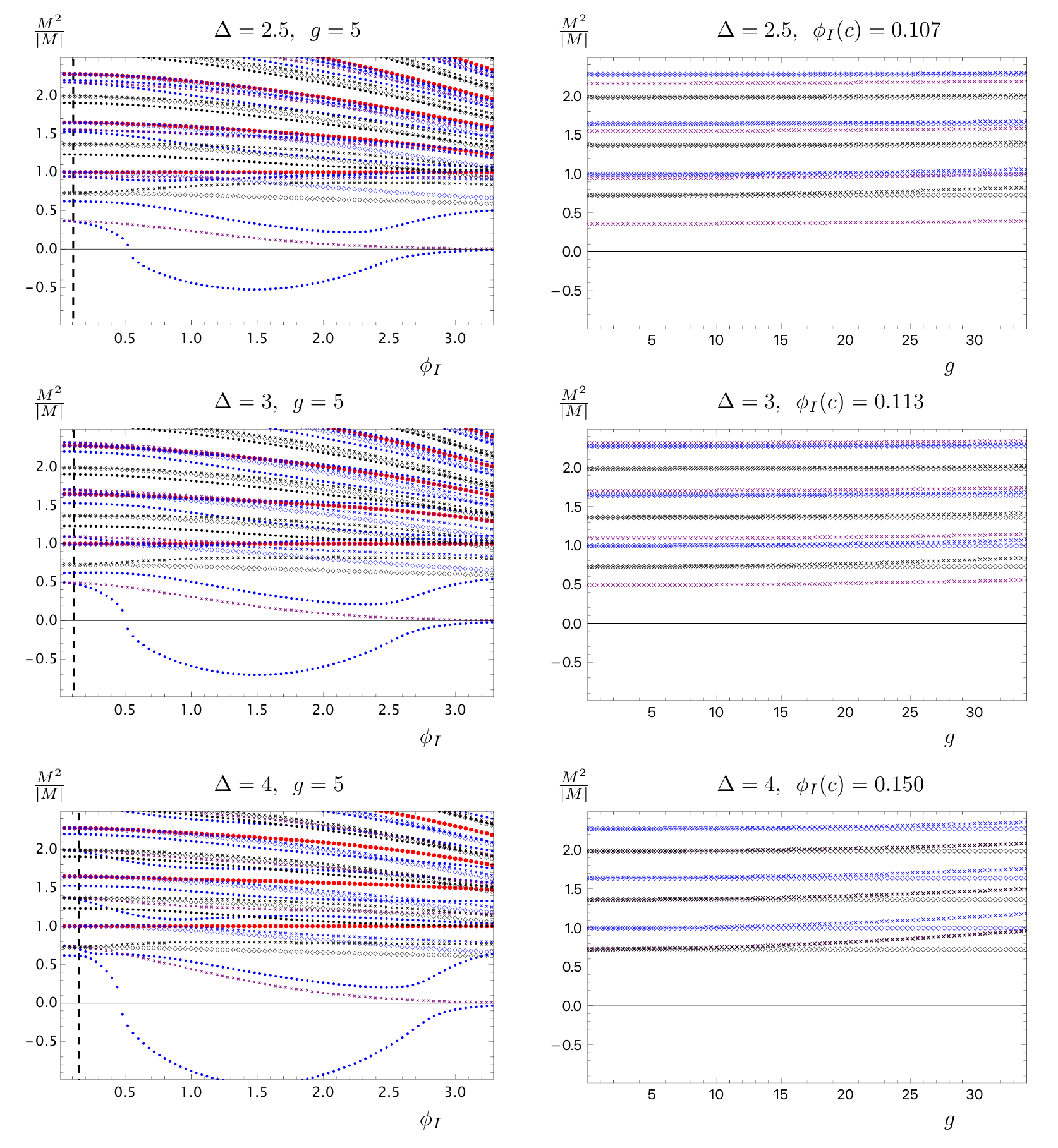}
\caption{Mass spectrum $\frac{M^2}{|M|}$ of fluctuations, computed for confining backgrounds, 
with various choices of $\Delta$, 
as a function of the IR parameter $\phi_I$ for $g=5$ (left),
and as a function of $g$ for $\phi_I=\phi_I(c)$ (right).
For each $\Delta$, we show the spectrum of scalar (blue), pseudo-scalar (purple),
vector (black), and tensor (red) states.  The values of the IR and UV cutoffs in the calculations are respectively given by $\rho_1 - \rho_o = 10^{-9}$ and  $\rho_2 - \rho_o = 5$, in all of the cases.
The different symbols refer to the quantum numbers in respect to the unbroken $SO(4)$ symmetry:
disks are used for singlets, and have already been reported in Ref.~\cite{Elander:2022ebt},
diamonds represent the $6$ of $SO(4)$, and crosses the $4$ of $SO(4)$.
All masses are normalised to the mass of the lightest spin-2 state.
Because the masses of the $SO(5)$ singlets do not depend on $g$, we do not repeat them in the right panels, 
which display only non-trivial $SO(5)$ multiplets.}
\label{Fig:mass2}
\end{center}
\end{figure}

 In order to discuss our new results and provide an interpretation for them,
we must first  pause and explain the physical meaning of the parameters $\phi_I$ and $g$.
The former is the parameter controlling the size of $SO(5)$ symmetry-breaking effects. Interestingly, despite the fact 
that $\phi$ obeys a second-order, non-linear differential equation, the requirement of regularity 
of the geometry at the end of the space  imposes a non-trivial
relation between the two free parameters that appear in a generic confining solution. For concreteness, we can think of them 
as the coefficients $\phi_J$ and $\phi_V$ appearing in the UV expansions, related to the explicit and spontaneous breaking of the symmetry on the field theory side.
The numerical study in Ref.~\cite{Elander:2022ebt} demonstrates that for 
$0<\Delta<5$, 
there is a critical value of $\phi_I$, denoted $\phi_I(c)$, such that when $\phi_I>\phi_I(c)$ 
there exists an alternative classical solution that has a lower value of the free energy,
for the same value of the source.
This result demonstrates the existence of a phase transition. 
We choose to display $\phi_I=\phi_I(c)$ in the plots because this choice minimises the mass of the lightest spin-0 state. 

The parameter $g$  controls the self-coupling of the bulk gauge fields. It is related to the coupling (and decay) of the composite vector mesons to two PNGBs, in the effective description of the dual field theory. But these statements require some more qualification, in view of the notational conventions we adopted. In the action, ${\cal S}_6$, we ignored a multiplicative factor of $2/\kappa^2$. In this paper, we are only solving classical background equations and  linearised equations for the  fluctuations around the chosen background solutions. In this process, $2/\kappa^2$  is just an overall factor that disappears from the final results. These classical results are exact if one takes  the limit $\kappa\rightarrow 0$ while holding fixed $g$ and $\phi_I$.
Close to the classical regime, perturbative corrections can be organised in loop diagrams, provided the coupling $g$ is not too strong. A naive estimate of the upper bound yields $\frac{3g^2\kappa^2}{256 \pi^3}\ll 1$~\cite{Ponton:2012bi}.\footnote{
The factor of $3$ in this expression is the second Casimir of the adjoint $C_2({\rm Adj})=3$ for $SO(5)$---it would be $C_2({\rm Adj})=N_c$ for $SU(N_c)$.}

    We show in Figs.~\ref{Fig:mass1} and~\ref{Fig:mass2} examples of mass
    spectra, for selected choices of $\Delta<5/2$ and 
 $\Delta\geq 5/2$, respectively.
Hence, for each representative choice  of $\Delta$,
 we produce one plot in which we fix $g=5$ and vary $\phi_I$, and a second plot in which we fix $\phi_I=\phi_I(c)$,
 and vary $g$. We show all the states of the system, differentiating them by colour, 
and shape of the markers  (for different $SO(4)$ representations).
We also reproduce the results for the $SO(5)$ singlet, for completeness of the presentation,
but also to set up their  physics implication. 
More examples of the numerical results are  presented in Appendix~\ref{Sec:more}.

For any values of $\Delta$, we find that the mass of the axial-vector states, transforming as $4$
 of $SO(4)$, is larger than that of the vectors, and the difference grows with $g$.
 Also, the mass of the lightest PNGBs grows with $g$.
When varying $\phi_I$ for $\Delta \lesssim  2$ and fixed $g$, the mass of the spin-0 states transforming 
as $4$ of $SO(4)$ grows with $\phi_I$. In field-theory terms, in this regime we are enhancing the effect of 
explicit symmetry breaking, compared to the spontaneous breaking, and there is no real sense in which these 
states are genuine  PNGBs, despite having the right quantum numbers.
But for $\Delta \geq 2.5$, we see that the mass of the lightest spin-0 states transforming as
 $4$ of $SO(4)$ can be made arbitrarily light, by choosing large values of $\phi_I$.
  Unfortunately though, the critical values of $\phi_I$ are rather small, and such large choices fall into
  the tachyonic part of the spectrum.
 The general conclusion of this exercise is that for all choices of $\Delta$ and $g$, if we restrict attention to the stable
 region of parameter space with $\phi_I\leq \phi_I(c)$, then the mass of the PNGBs
  shows no indications of being  suppressed.

\begin{figure}[th]
\begin{center}
\includegraphics[width=16cm]{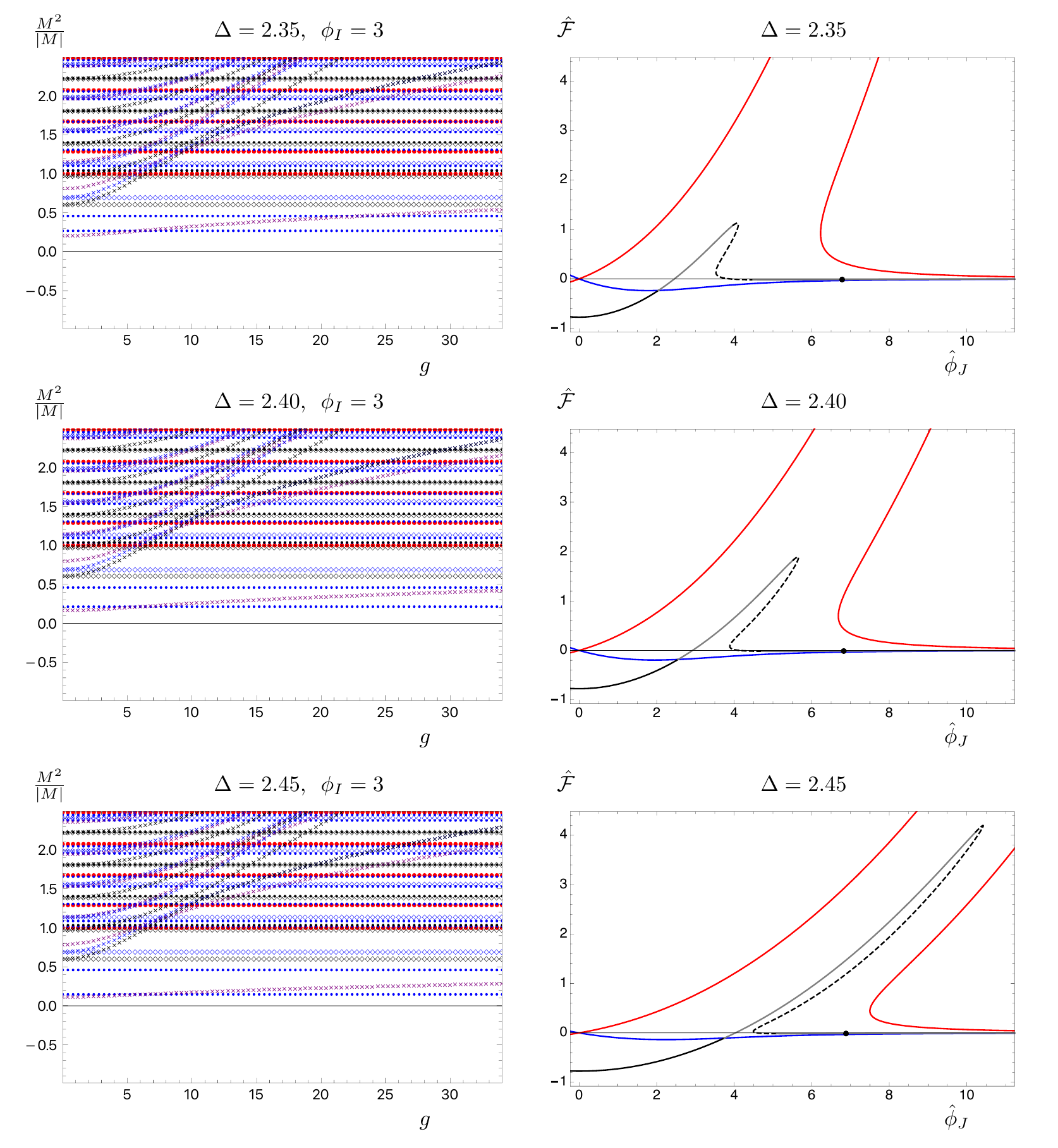}
\caption{Mass spectrum $\frac{M^2}{|M|}$ of fluctuations, computed for confining backgrounds, 
for $\Delta=2.35$ (top), $\Delta=2.40$ (middle), and  $\Delta=2.45$ (bottom), and $\phi_I=3$
as a function of $g$ (left),
and free energy $\hat{\cal F}$ a function of the normalised source $\hat{\phi}_J$ (right). 
For the spectrum,  scalar (blue), pseudo-scalar (purple),
vector (black), and tensor (red)  states are displayed as disks for $SO(4)$ singlets,
as diamonds to represent the $6$ of $SO(4)$, and crosses for the $4$ of $SO(4)$. 
The values of the IR and UV cutoffs in the calculations are respectively given by $\rho_1 - \rho_o = 10^{-9}$ and  $\rho_2 - \rho_o = 5$.
All masses are normalised to the mass of the lightest spin-2 state.
The definition of (normalised) free energy $\hat{\cal F}$ and (normalised) source $\hat{\phi}_J$
can be found in  Ref.~\cite{Elander:2022ebt}.
The black (stable), gray (metastable), and dashed black (tachyonic) regions of the curve refers to the confining solutions of interest, while red and blue curves refer to singular solutions. The black dots on the right  panels denote the solutions with $\phi_I=3$.}
\label{Fig:detail}
\end{center}
\end{figure}

 Interestingly, we find something new when we focus attention on the case where $2\lesssim \Delta<2.5$.  Ref.~\cite{Elander:2022ebt} found the existence of a metastable region of parameter space with 
  large $\phi_I$, in which the lightest scalar is a dilaton.
 Here, we find that also the PNGBs, transforming as a $4$ of $SO(4)$, are  light in this region of parameter space,
 their masses being suppressed in respect to the scale of all other bound states. 
 This can be seen in the bottom-left panel of Fig.~\ref{Fig:mass1}.
 
 To demonstrate that the mass of these two states can be dialled to 
  be arbitrarily small, compared to the typical mass scale of 
 all other bound states,  represented by the mass of the spin-2 particles,
 in Fig.~\ref{Fig:detail}, we display some more information about the choices $\Delta=2.35$, 
 $\Delta=2.40$,  and $\Delta=2.45$. 
 We show in the left panels of the figure 
 the dependence on $g$ of the spectrum, for a  choice of $\phi_I=3$, large enough 
to fall in the portion of parameter space that contains a light dilaton together with a light 
 set of PNGBs transforming as the $4$ of $SO(4)$.

 While Ref.~\cite{Elander:2022ebt}, for such large values of $\phi_I$, found that the confining background solutions 
are metastable, we produce here three expanded and detailed plots showing
 that the free energy is almost degenerate with another branch of solutions.
The plots on the right panels of Fig.~\ref{Fig:detail} show the free energy  $\hat{\cal F}$ 
computed using holographic renormalisation,
as a function of the source $\hat{\phi}_J$, and normalised appropriately. 
The plots are expanded versions of those in Ref.~\cite{Elander:2022ebt}, and 
for these choices are obtained with  the following relations:
\beqs
{\cal F} &=& -\frac{1}{40} e^{4A_U-\chi_U}\left(16\Delta\left(\frac{5}{2}-\Delta\right)\phi_J\phi_V-75\chi_5\right)\,,\\
\Lambda^{-1}&\equiv&\int_{\rho_o}^{\infty} \dd \rho\,e^{\chi(\rho)-A(\rho)}\,,
\eeqs
and the rescaling $\hat{\cal F}\equiv {\cal F}/\Lambda^5$, $\hat{\phi}_J\equiv \phi_J/\Lambda^{\Delta}$. 
We do not repeat here details, except for specifying that in the plots  the choice $\phi_I=3$ is equivalent to confining solutions with $\hat{\phi}_J=6.78$ for $\Delta=2.35$,
$\hat{\phi}_J=6.83$ for $\Delta=2.40$,
and
 $\hat{\phi}_J=6.88$
for $\Delta=2.45$. 
The reason why we show these plots is that, besides the confining solutions, the analysis of the free energy 
carries over also for (singular) solutions respecting five-dimensional Poincar\'e invariance,
and we can show a graphical comparison. In particular one can explicitly see that 
for $\phi_I=3$, in all three examples reported here, the confining solutions do not minimise the
free energy.
In the limit of large $\phi_I$, the metastable solutions might be long lived. Whether there are regions of parameter space that allow for the construction of a viable CHM, relying on the existence of a long-lived metastable vacuum, is an important question that would require a dedicated study.

While for small $\phi_I$ one could argue that the choice of potential adopted in this paper is as good as any,
because it can be obtained as a power-expansion of more complicated potentials in the regime of small $\phi$,
one expects model-dependence to affect the large $\phi_I$ region.
Top-down models are not affected by this limitation. For example, Fig.~7 of Ref.~\cite{Elander:2021kxk},
shows the spectrum of a top-down model with $SO(5)$ symmetry breaking to $SO(4)$. 
The results at large $\phi_I$ qualitatively resembles,
for large values of the VEV, the case $\Delta\geq 5/2$ of this paper, as
for arbitrarily  large values of $\phi_I$ the PNGBs become arbitrarily light,
but also other states, including a tachyon, persist and their masses appear to be suppressed
compared to the typical scale of the other bound state masses.
We do not know if top-down models showing the feature we uncovered here exist,
namely in which both dilaton and PNGBs are parametrically light, but there is no tachyon.
Nevertheless, this is a new, unexpected result, which might have important phenomenological implications, 
that are worth studying in the future.

\section{Outlook}
\label{Sec:outlook}

In this paper, we studied the spectrum of bound states carrying $SO(5)$ quantum numbers, 
in a strongly-coupled, confining  field theory modelled by its higher-dimensional, weakly-coupled gravity dual. 
We paid particular attention to the states that have the correct quantum numbers to be
identified as PNGBs,
 as the  $SO(5)$ global symmetry of the field theory is broken both explicitly and spontaneously to its $SO(4)$ subgroup.
We studied the spectrum as a function of three parameters. $\Delta$ is the parameter that, for $\Delta>5/2$, 
is interpreted in the field theory as 
the dimension of the 
scalar operator controlling $SO(5)$ breaking, 
and for $\Delta<5/2$ as the dimension of the coupling of the operator.
$\phi_I$ is the parameter controlling the size of the symmetry breaking.
And $g$ controls the self-coupling of vector fields, as well as their coupling to the PNGBs.

The main results of our analysis are two-fold. First, we showed that if we
restrict our attention to the region of parameter space in which the confining solutions are stable,
as identified in Ref.~\cite{Elander:2022ebt}, then neither the scalar $SO(5)$ singlet nor the $SO(4)$ multiplets
are parametrically light, for any values of $\phi_I\leq\phi_I(c)$, $\Delta$, and $g$ that we considered. Second, we identified a metastable region of parameter space with $2\lesssim \Delta< 2.5$ and large $\phi_I$,
for which both the scalar $SO(5)$ singlet and the lightest spin-0 states transforming as $4$ of $SO(4)$
become arbitrarily light when approaching $\Delta\rightarrow 5/2$.
 In this case, the former is a dilaton, and the latter is a multiplet of PNGBs.
The existence of this region of parameter space, and the fact that both types of particles are parametrically light,
are both new and unexpected results, deserving further investigation.

This is the first step towards the construction of a composite Higgs model, in which the Higgs fields emerge as the PNGBs of a new strongly-coupled theory. The next step requires to couple the system to the SM gauge fields, and  to study vacuum alignment in the theory, as a function of  the strength of additional symmetry-breaking parameters, which in the gravity theory correspond to boundary-localised terms. Whether or not this will allow to explore other, enlarged regions of parameter space as viable for CHM model-building is not known, as is not known whether the presence of the dilaton in the metastable region of parameter space has phenomenologically relevant implications. These interesting questions will be addressed in future research.

\begin{acknowledgments}

The work of AF has been supported by the STFC Consolidated Grant No. ST/V507143/1.

The work of  MP has been supported in part by the STFC 
Consolidated Grants No. ST/P00055X/1 and No. ST/T000813/1.
  MP received funding from
the European Research Council (ERC) under the European
Union’s Horizon 2020 research and innovation program
under Grant Agreement No.~813942. 

\vspace{0.5cm}

{\bf Research Data Access Statement}---The data generated for this manuscript can be downloaded from
Ref.~\cite{Data}.

\vspace{0.5cm}

{\bf Open Access Statement}---For the purpose of open access, the authors have applied a Creative Commons 
Attribution (CC BY) licence  to any Author Accepted Manuscript version arising.

\end{acknowledgments}


\appendix

\allowdisplaybreaks

\section{Basis of $SO(5)$ generators}
\label{Sec:so5}

For concreteness, we present here an example of a basis of $SO(5)$ generators,
which we chose so that the first four generators $t^{\hat{A}}$, with $\hat{A}=1\,,\,\cdots\,,\,4$, span the
the coset $SO(5)/SO(4)$, with the conventions in Eq.~(\ref{eq:Xdecomp}), 
while  the unbroken $SO(4)$ is generated by 
$t^{\bar{A}}$, with $\bar{A}=5\,,\,\cdots\,,\,10$.
\beqs
t^1 &=& \frac{i}{2} \left(
\begin{array}{ccccc}
0 & 0 & 0 & 0 &-1 \\
0 & 0 & 0 & 0 & 0 \\
0 & 0 & 0 & 0 & 0 \\
0 & 0 & 0 & 0 & 0 \\
1 & 0 & 0 & 0 & 0
\end{array}\right) \,, \quad
t^2 = \frac{i}{2} \left(\begin{array}{ccccc}
0 & 0 & 0 & 0 & 0 \\
0 & 0 & 0 & 0 & -1 \\
0 & 0 & 0 & 0 & 0 \\
0 & 0 & 0 & 0 & 0 \\
0 & 1 & 0 & 0 & 0
\end{array}\right) \,, \quad
t^3 = \frac{i}{2} \left(\begin{array}{ccccc}
0 & 0 & 0 & 0 & 0 \\
0 & 0 & 0 & 0 & 0 \\
0 & 0 & 0 & 0 & -1 \\
0 & 0 & 0 & 0 & 0 \\
0 & 0 & 1 & 0 & 0
\end{array}\right) \,, \quad
t^4 = \frac{i}{2} \left(\begin{array}{ccccc}
0 & 0 & 0 & 0 & 0 \\
0 & 0 & 0 & 0 & 0 \\
0 & 0 & 0 & 0 & 0 \\
0 & 0 & 0 & 0 & -1 \\
0 & 0 & 0 & 1 & 0
\end{array}\right) \,, \nonumber \\
t^5 &=& \frac{i}{2} \left(\begin{array}{ccccc}
0 & 0 & 0 & -1 & 0 \\
0 & 0 & 0 & 0 & 0 \\
0 & 0 & 0 & 0 & 0 \\
1 & 0 & 0 & 0 & 0 \\
0 & 0 & 0 & 0 & 0
\end{array}\right) \,, \quad
t^6 = \frac{i}{2} \left(\begin{array}{ccccc}
0 & 0 & 0 & 0 & 0 \\
0 & 0 & 0 & -1 & 0 \\
0 & 0 & 0 & 0 & 0 \\
0 & 1 & 0 & 0 & 0 \\
0 & 0 & 0 & 0 & 0
\end{array}\right) \,, \quad
t^7 = \frac{i}{2} \left(\begin{array}{ccccc}
0 & 0 & 0 & 0 & 0 \\
0 & 0 & 0 & 0 & 0 \\
0 & 0 & 0 & -1 & 0 \\
0 & 0 & 1 & 0 & 0 \\
0 & 0 & 0 & 0 & 0
\end{array}\right) \,, \nonumber \\
t^8 &=& \frac{i}{2} \left(\begin{array}{ccccc}
0 & 0 & -1 & 0 & 0 \\
0 & 0 & 0 & 0 & 0 \\
1 & 0 & 0 & 0 & 0 \\
0 & 0 & 0 & 0 & 0 \\
0 & 0 & 0 & 0 & 0
\end{array}\right) \,, \quad
t^9 = \frac{i}{2} \left(\begin{array}{ccccc}
0 & 0 & 0 & 0 & 0 \\
0 & 0 & -1 & 0 & 0 \\
0 & 1 & 0 & 0 & 0 \\
0 & 0 & 0 & 0 & 0 \\
0 & 0 & 0 & 0 & 0
\end{array}\right) \,, \quad
t^{10} = \frac{i}{2} \left(\begin{array}{ccccc}
0 & -1 & 0 & 0 & 0 \\
1 & 0 & 0 & 0 & 0 \\
0 & 0 & 0 & 0 & 0 \\
0 & 0 & 0 & 0 & 0 \\
0 & 0 & 0 & 0 & 0
\end{array}\right) \,. \quad
\eeqs

\section{Asymptotic expansions of the fluctuations}
\label{sec:IRUVexpansions}

The linearised equations governing the dynamics of the small fluctuations 
around the classical solutions are subject to boundary conditions that, as 
explained in the body of the paper, can be implemented by matching to the asymptotic expansion of the solutions,
in a way that resembles the process of improvement in lattice field theory.
It is hence useful to report here such asymptotic expansions.
We find it convenient to  include also 
the $SO(5)$ singlets, together with the $SO(4)$ multiplets.

\subsection{IR expansions}

We start from the IR expansion of the fluctuations.
For convenience, we put $\rho_o = 0$ and $A_I = 0$ in this subsection,\footnote{The dependence on 
$\rho_o$ and $A_I$ can be reinstated by making the substitutions $\rho \rightarrow \rho - \rho_o$ and 
$q^2 \rightarrow e^{-2A_I} q^2$ in the expressions.} while setting $\chi_I = 0$ in 
order to avoid a conical singularity.
We then expand the solutions of the linearised equations in powers of small $\rho$.
We can write the expansion for a general value of $\Delta$.

For the scalar fluctuations, we find
\begin{align}
\mathfrak a^1 =&\ \mathfrak a^1_{I,0}+\mathfrak a^1_{I,l} \log (\rho )+\frac{1}{4} \rho ^2 \bigg[ -\frac{1}{4} \Delta  \left(\mathfrak a^1_{I,0} \left(\Delta  \left(15 \phi_I^2-4\right)+20\right)+6 \phi_I (\mathfrak a^2_{I,0}-\mathfrak a^2_{I,l}) \left(\Delta  \left(5 \phi_I^2-4\right)+20\right)\right)
\nonumber \\ &
+q^2 (\mathfrak a^1_{I,0}-\mathfrak a^1_{I,l}) -\frac{1}{48} \mathfrak a^1_{I,l} \left(\Delta  \left(25 \Delta  \phi_I^4+20 (10-11 \Delta ) \phi_I^2+48 (\Delta -5)\right)+400\right)
\nonumber \\ &
+\log (\rho ) \left(\mathfrak a^1_{I,l} \left(-\frac{15 \Delta ^2 \phi_I^2}{4}+(\Delta -5) \Delta +q^2\right)-\frac{3}{2} \mathfrak a^2_{I,l} \Delta  \phi_I \left(\Delta  \left(5 \phi_I^2-4\right)+20\right)\right) \bigg]+\mathcal O \left(\rho ^4\right) \,, \\ 
\mathfrak a^2 =&\ \mathfrak a^2_{I,0}+\mathfrak a^2_{I,l} \log (\rho )+\frac{1}{4} \rho ^2 \bigg[-\frac{1}{4} \Delta  \phi_I (\mathfrak a^1_{I,0}-\mathfrak a^1_{I,l}) \left(\Delta  \left(5 \phi_I^2-4\right)+20\right) +q^2 (\mathfrak a^2_{I,0}-\mathfrak a^2_{I,l})
\nonumber \\ &
-\frac{3}{8} \mathfrak a^2_{I,0} \left(\Delta  \phi_I^2 \left(\Delta  \left(5 \phi_I^2-8\right)+40\right)+80\right)+\frac{13}{48} \mathfrak a^2_{I,l} \left(\Delta  \phi_I^2 \left(\Delta  \left(5 \phi_I^2-8\right)+40\right)+80\right)
\nonumber \\ &
+\log (\rho ) \left(-\frac{5}{4} \mathfrak a^1_{I,l} \Delta ^2 \phi_I^3+\mathfrak a^1_{I,l} (\Delta -5) \Delta  \phi_I+\mathfrak a^2_{I,l} \left(-\frac{15}{8} \Delta ^2 \phi_I^4+3 (\Delta -5) \Delta  \phi_I^2+q^2-30\right)\right)\bigg]+\mathcal O \left(\rho ^4\right) \,, \\
\mathfrak a^3 =&\ \mathfrak a^3_{I,0}+\rho ^2 \left(\frac{1}{2} \mathfrak a^3_{I,0} q^2 \log (\rho )+\mathfrak a^3_{I,2}\right)+\mathcal O \left(\rho ^4\right) \,, \\ 
\mathfrak a^4 =&\ \mathfrak a^4_{I,0}+\rho ^2 \left(\frac{1}{2} \mathfrak a^4_{I,0} \left( q^2+ \frac{g^2\phi_I^2}{4} \right) \log (\rho )+\mathfrak a^4_{I,2}\right)+\mathcal O \left(\rho ^4\right) \,.
\end{align}

For the pseudo-scalar fluctuations, we find
\beq
\mathfrak p = \mathfrak p_{I,0}+\rho ^2 \left[\mathfrak p_{I,2} + \frac{1}{2} \mathfrak p_{I,0} \left(q^2+\frac{g^2}{4}\phi_I^2\right) \log (\rho )\right]+\mathcal O \left(\rho ^4\right) \,.
\eeq

For the vector fluctuations, we find
\begin{align}
\mathfrak v^1 =&\ \mathfrak v^1_{I,-2} \rho ^{-2}+\frac{1}{2} q^2 \mathfrak v^1_{I,-2} \log (\rho )+\mathfrak v^1_{I,0} +\frac{1}{12288} \rho ^2 \Big[1536 q^2 \mathfrak v^1_{I,0}+80 \Delta ^2 \mathfrak v^1_{I,-2} \phi_I^4 \left(2 \left(8 \Delta ^2-50 \Delta +75\right)-3 q^2\right)
\nonumber \\ &
+128 (\Delta -5) \Delta  \mathfrak v^1_{I,-2} \phi_I^2 \left(-3 (\Delta -5) \Delta +3 q^2-50\right)-64 \left(9 q^4+60 q^2-500\right) \mathfrak v^1_{I,-2}+125 \Delta ^4 \mathfrak v^1_{I,-2} \phi_I^8
\nonumber \\ &
-1000 (\Delta -2) \Delta ^3 \mathfrak v^1_{I,-2} \phi_I^6+ 768 q^4 \mathfrak v^1_{I,-2} \log (\rho )\Big]+\mathcal O \left(\rho ^4\right) \,, \\ 
\mathfrak v^2 =&\ \mathfrak v^2_{I,0}+\mathfrak v^2_{I,l} \log (\rho )+\frac{1}{96} \rho ^2 \Big[24 q^2 (\mathfrak v^2_{I,0}-\mathfrak v^2_{I,l})+\mathfrak v^2_{I,l} \left(-5 \Delta ^2 \phi_I^4+8 (\Delta -5) \Delta  \phi_I^2-80\right)
\nonumber \\ &
+24 q^2 \mathfrak v^2_{I,l} \log (\rho )\Big]+\mathcal O \left(\rho ^4\right) \,, \\
\mathfrak v^3 =&\ \mathfrak v^3_{I,0}+\mathfrak v^3_{I,l} \log (\rho )+\frac{1}{96} \rho ^2 \Big[\left(24 q^2+6 g^2 \phi_I^2\right) \mathfrak v^3_{I,0}+\left(-80-24 q^2-6 g^2 \phi_I^2-40 \Delta  \phi_I^2+\Delta ^2 \left(8 \phi_I^2-5 \phi_I^4\right)\right) \mathfrak v^3_{I,l}\nonumber \\ &+\left(24 q^2+6 g^2 \phi_I^2\right) \log (\rho ) \mathfrak v^3_{I,l}
   \Big]+\mathcal O \left(\rho ^4\right) \,.
\end{align}

For the tensor fluctuations, we find
\beq
\mathfrak e = \mathfrak e_{I,0}+\mathfrak e_{I,l} \log (\rho )+\frac{1}{192} \rho ^2 \Big[48 q^2 (\mathfrak e_{I,0}-\mathfrak e_{I,l})-25 \Delta ^2 \mathfrak e_{I,l} \phi_I^4+40 (\Delta -5) \Delta  \mathfrak e_{I,l} \phi_I^2-400 \mathfrak e_{I,l}+48 \mathfrak e_{I,l} q^2 \log (\rho )\Big]+\mathcal O \left(\rho ^4\right) \,.
\eeq

\subsection{UV expansions}

The expansions for large $\rho$, in the UV regime of the dual field-theory interpretation, 
depend non-trivially on the parameter $\Delta$. For illustration purposes,
in this subsection  we set $\Delta = 3$, and $A_U = 0 = \chi_U$.\footnote{The dependence on $\chi_U$ and $A_U$ can be reinstated by making the substitution $q^2 \rightarrow e^{2\chi_U-2A_U} q^2$ in the expressions.} We write the expansions in terms of $z \equiv e^{-\rho}$.

For the scalar fluctuations, we find
\begin{align}
\mathfrak a^1 =&\ \mathfrak a^1_2 z^2+\mathfrak a^1_3 z^3+\frac{1}{2} \mathfrak a^1_2 q^2 z^4+\frac{1}{6} \mathfrak a^1_3 q^2 z^5+\frac{1}{48} \mathfrak a^1_2 \left(2 q^4-99 \phi_J^2\right) z^6 +\mathcal O \left(z^7\right) \,, \\ 
\mathfrak a^2 =&\ \mathfrak a^2_0-\frac{1}{6} \mathfrak a^2_0 q^2 z^2 +\frac{1}{24} \mathfrak a^2_0 q^4 z^4+\mathfrak a^2_5 z^5+\frac{1}{144} \mathfrak a^2_0 q^2 \left(q^4-14 \phi_J^2\right) z^6 +\mathcal O \left(z^7\right)\,, \\
\mathfrak a^3 =&\ \mathfrak a^3_0-\frac{1}{2} \mathfrak a^3_0 q^2 z^2 +\mathfrak a^3_3 z^3-\frac{1}{8} \mathfrak a^3_0 q^4 z^4 +\frac{1}{10} \mathfrak a^3_3 q^2 z^5-\frac{1}{144} \mathfrak a^3_0 q^2 \left(q^4+10 \phi_J^2\right) z^6 +\mathcal O \left(z^7\right) \,, \\ 
\mathfrak a^4 =&\ \mathfrak a^4_0-\frac{1}{2} \mathfrak a^4_0 q^2 z^2 +\mathfrak a^4_3 z^3+\frac{1}{16} \mathfrak a^4_0 \left(g^2\phi_J^2-2 q^4\right) z^4 +\frac{1}{20} \left(\mathfrak a^4_0 g^2\phi_J \phi_V+2 \mathfrak a^4_3 q^2\right) z^5
\nonumber \\ &
-\frac{1}{288} \mathfrak a^4_0 \left(2 q^6+(20+g^2) q^2 \phi_J^2-4g^2\phi_V^2\right) z^6 +\mathcal O \left(z^7\right) \,.
\end{align}

For the pseudo-scalar fluctuations, we find
\begin{align}
\mathfrak p =&\ \mathfrak p_0+\mathfrak p_1 z+ \left(\frac{\mathfrak p_0 q^2}{2}+\frac{\mathfrak p_1 \phi_V}{\phi_J}\right) z^2 +\frac{2 \mathfrak p_0 q^2 \phi_J \phi_V+\mathfrak p_1 q^2 \phi_J^2+2 \mathfrak p_1 \phi_V^2}{6 \phi_J^2} z^3 
\nonumber \\ &
+\frac{ \mathfrak p_0 \left(q^4 \phi_J +\frac{g^2 \phi_J ^3}{2}\right)+4 \mathfrak p_1 q^2 \phi_V }{24 \phi_J }z^4
+\mathcal O \left(z^5\right) \,.
\end{align}

For the vector fluctuations, we find
\begin{align}
\mathfrak v^1 =&\ \mathfrak v^1_0-\frac{1}{6} q^2 \mathfrak v^1_0 z^2 +\frac{1}{24} q^4 \mathfrak v^1_0 z^4+\mathfrak v^1_5 z^5+\frac{1}{144} q^2 \mathfrak v^1_0 \left(q^4-14 \phi_J^2\right) z^6
\nonumber \\ &
+\frac{1}{70} q^2 (70 \mathfrak v^1_0 \chi_5-2 \mathfrak v^1_0 \phi_J \phi_V+5 \mathfrak v^1_5) z^7 +\mathcal O \left(z^8\right) \,, \\ 
\mathfrak v^2 =&\ \mathfrak v^2_0-\frac{1}{2} q^2 \mathfrak v^2_0 z^2 +\mathfrak v^2_3 z^3-\frac{1}{8} q^4 \mathfrak v^2_0 z^4 +\frac{1}{10} q^2 \mathfrak v^2_3 z^5-\frac{1}{144} q^2 \mathfrak v^2_0 \left(q^4+10 \phi_J^2\right) z^6
\nonumber \\ &
+\frac{1}{1400} \left(5(q^4+45\phi_J^2) \mathfrak v^2_3+6 q^2 \mathfrak v^2_0 (75 \chi_5-26 \phi_J \phi_V)\right) z^7 +\mathcal O \left(z^8\right) \,, \\
\mathfrak v^3 =&\ \mathfrak v^3_0-\frac{1}{2} q^2 \mathfrak v^3_0 z^2 +\mathfrak v^3_3 z^3-\frac{1}{8} \mathfrak v^3_0 \left(q^4- \frac{g^2}{2}\phi_J^2\right) z^4 +\frac{1}{10} \left(q^2 \mathfrak v^3_3+\frac{g^2}{2} \mathfrak v^3_0 \phi_J \phi_V\right) z^5 +\mathcal O \left(z^6\right) \,.
\end{align}

For the tensor fluctuations, we find
\beq
\mathfrak e = \mathfrak e_0- \frac{1}{6} \mathfrak e_0 q^2 z^2 +\frac{1}{24} \mathfrak e_0 q^4 z^4+\mathfrak e_5 z^5+\mathcal O \left(z^6\right) \,.
\eeq

The choice $\Delta=3$ yields a particularly simple expansion in powers of $z$. In the process of carrying out the numerical calculations for this paper, we computed the UV expansions for all values of $\Delta$ for which we plot the spectrum. We do not report all of these expansions here, but we notice that for special choices of $\Delta$ the  formal expansion changes, to include also logarithmic terms in the form $z^n \log^m (z)$.

\section{More mass spectra}
\label{Sec:more}

\begin{figure}[th]
\begin{center}
\includegraphics[width=16cm]{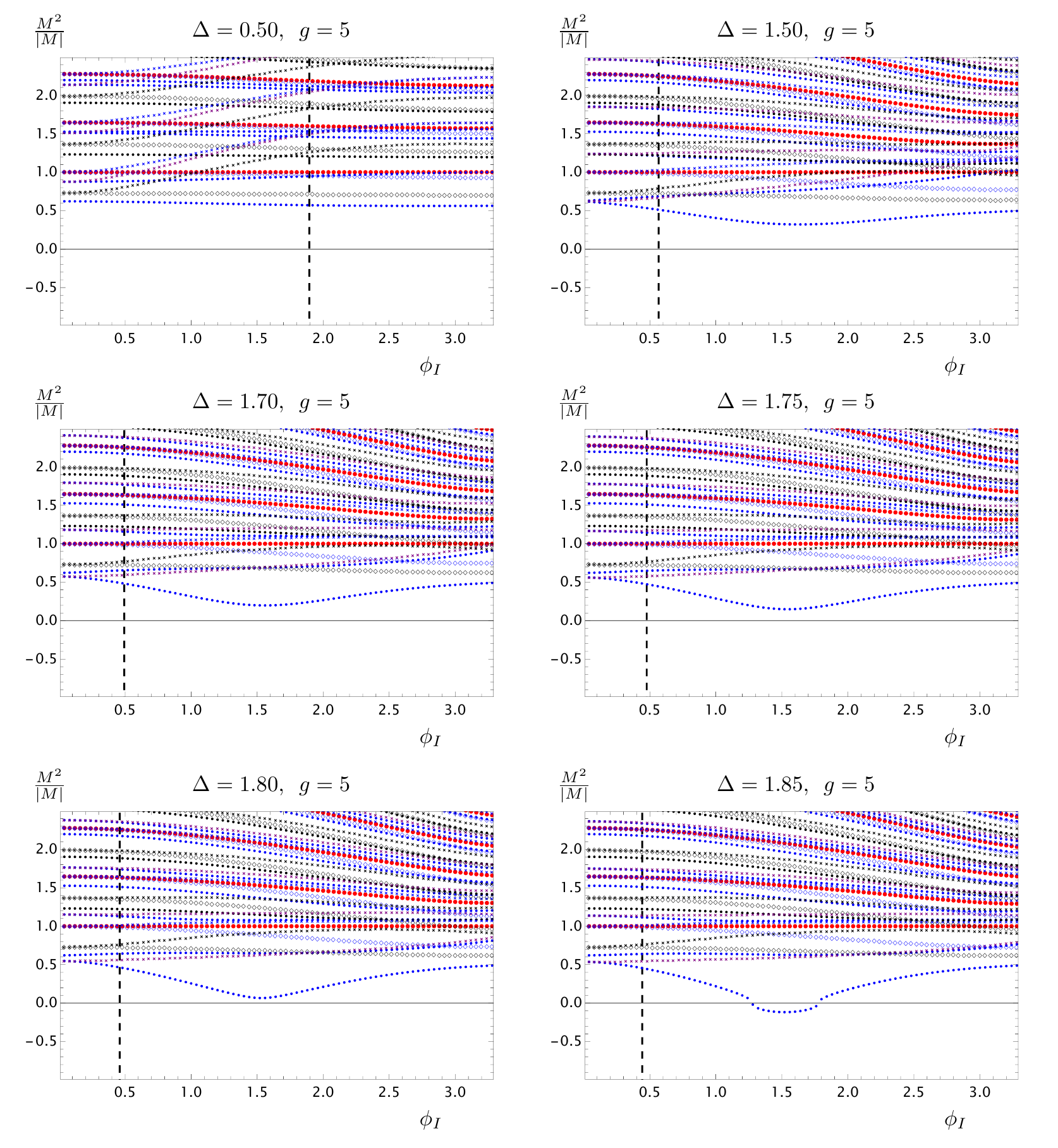}
\caption{Mass spectrum $\frac{M^2}{|M|}$ of fluctuations, computed for confining backgrounds, 
with various choices of $\Delta$, 
as a function of the IR parameter $\phi_I$ for $g=5$.
For each $\Delta$, we show the spectrum of scalar (blue), pseudo-scalar (purple),
vector (black), and tensor (red) states.  The values of the IR and UV cutoffs in the calculations are respectively given by $\rho_1 - \rho_o = 10^{-9}$ and  $\rho_2 - \rho_o = 5$, in all of the cases.
The different symbols refer to the quantum numbers in respect to the unbroken $SO(4)$ symmetry:
disks are used for singlets, and have already been reported in Ref.~\cite{Elander:2022ebt},
diamonds represent the $6$ of $SO(4)$, and crosses the $4$ of $SO(4)$.
All masses are normalised to the mass of the lightest spin-2 state.}
\label{Fig:mass3}
\end{center}
\end{figure}

\begin{figure}[th]
\begin{center}
\vspace{-0.5cm}
\includegraphics[width=16cm]{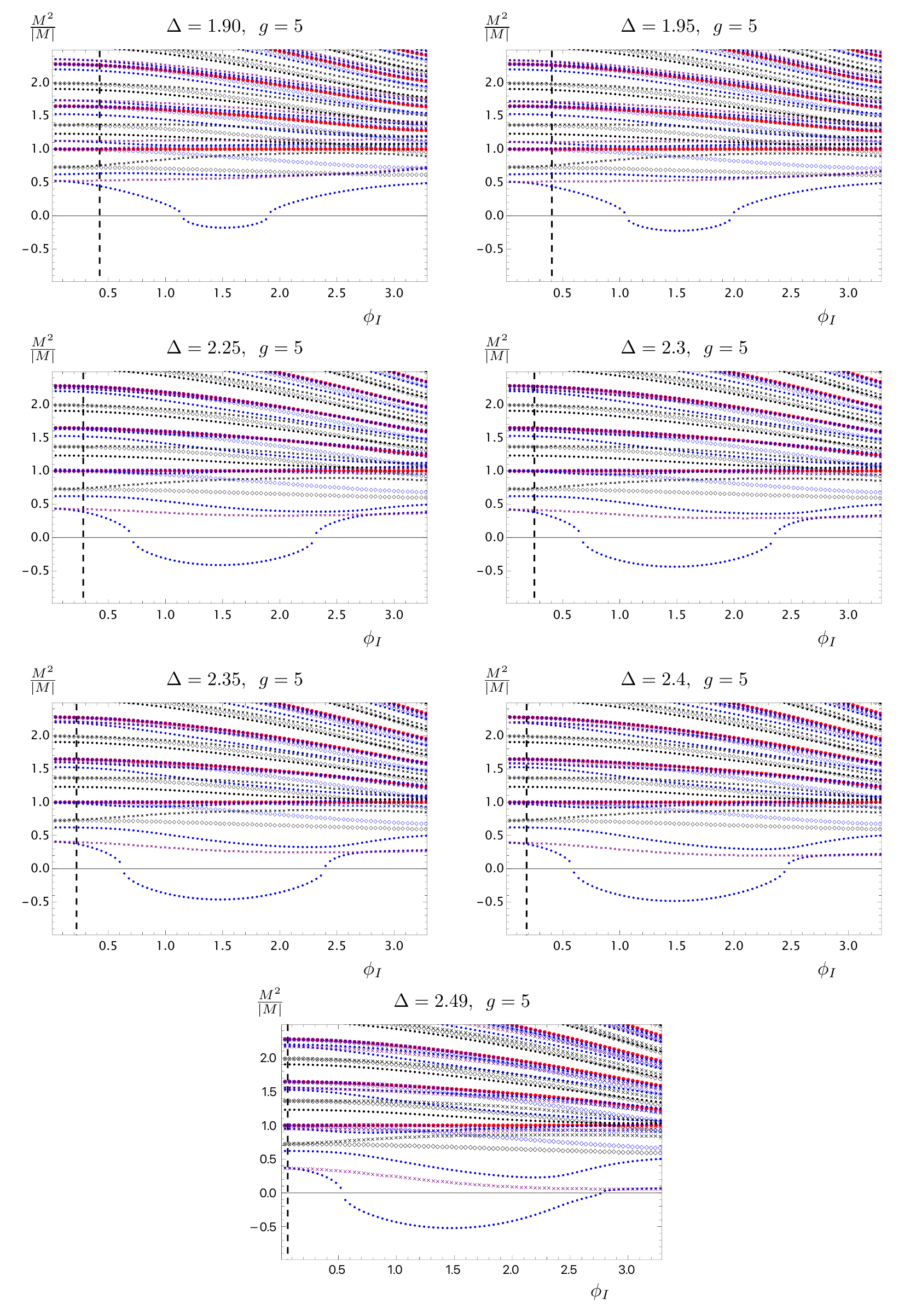}
\vspace{-0.5cm}
\caption{Mass spectrum $\frac{M^2}{|M|}$ of fluctuations, computed for confining backgrounds, 
with various choices of $\Delta$, 
as a function of the IR parameter $\phi_I$ for $g=5$.
For each $\Delta$, we show the spectrum of scalar (blue), pseudo-scalar (purple),
vector (black), and tensor (red) states.  The values of the IR and UV cutoffs in the calculations are respectively given by $\rho_1 - \rho_o = 10^{-9}$ and  $\rho_2 - \rho_o = 5$, in all of the cases.
The different symbols refer to the quantum numbers in respect to the unbroken $SO(4)$ symmetry:
disks are used for singlets, and have already been reported in Ref.~\cite{Elander:2022ebt},
diamonds represent the $6$ of $SO(4)$, and crosses the $4$ of $SO(4)$.
All masses are normalised to the mass of the lightest spin-2 state.}
\label{Fig:mass4}
\end{center}
\end{figure}

\begin{figure}[th]
\begin{center}
\vspace{-0.5cm}
\includegraphics[width=16cm]{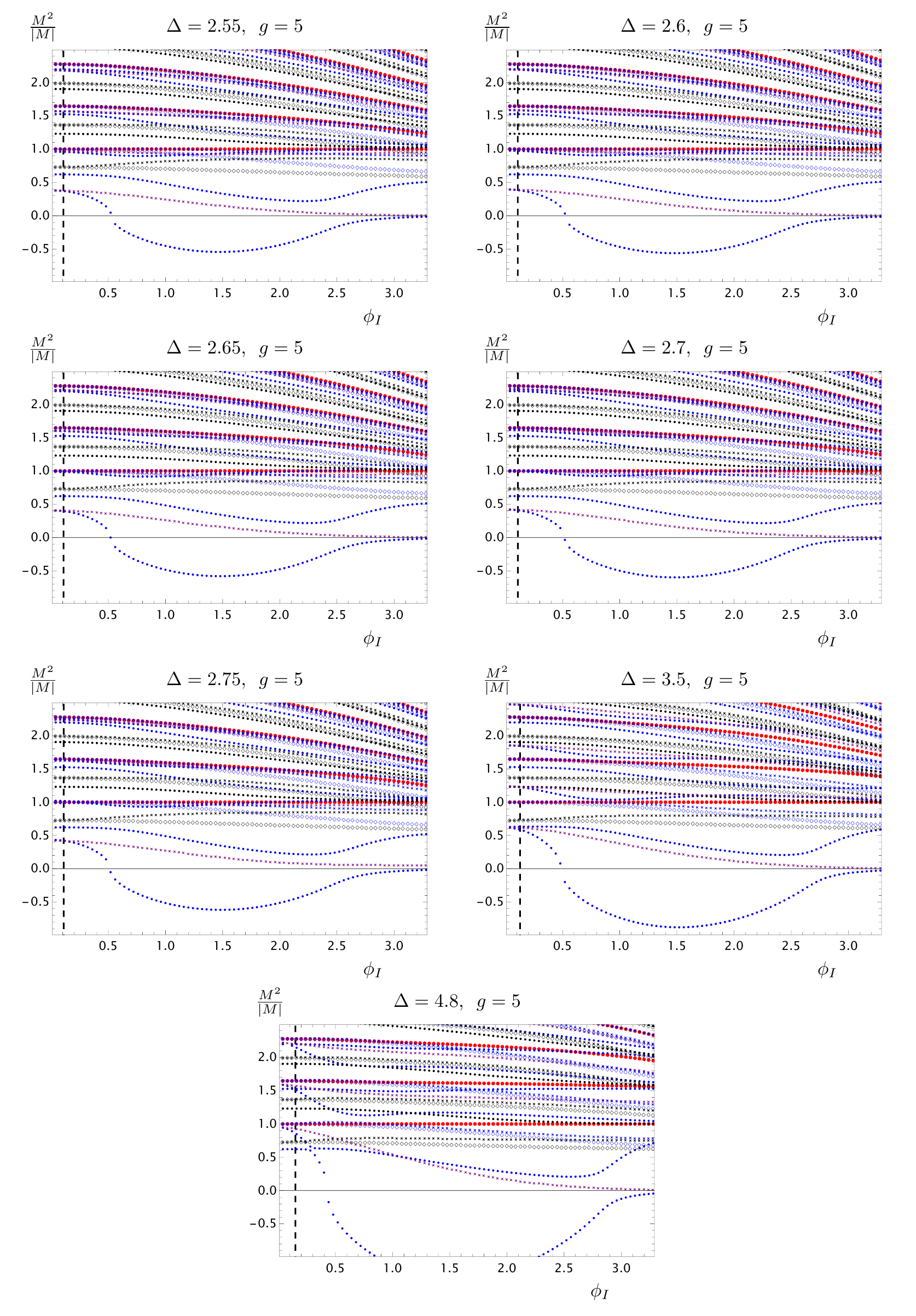}
\vspace{-0.5cm}
\caption{Mass spectrum $\frac{M^2}{|M|}$ of fluctuations, computed for confining backgrounds, 
with various choices of $\Delta$, 
as a function of the IR parameter $\phi_I$ for $g=5$.
For each $\Delta$, we show the spectrum of scalar (blue), pseudo-scalar (purple),
vector (black), and tensor (red) states.  The values of the IR and UV cutoffs in the calculations are respectively given by $\rho_1 - \rho_o = 10^{-9}$ and  $\rho_2 - \rho_o = 5$, in all of the cases.
The different symbols refer to the quantum numbers in respect to the unbroken $SO(4)$ symmetry:
disks are used for singlets, and have already been reported in Ref.~\cite{Elander:2022ebt},
diamonds represent the $6$ of $SO(4)$, and crosses the $4$ of $SO(4)$.
All masses are normalised to the mass of the lightest spin-2 state.}
\label{Fig:mass5}
\end{center}
\end{figure}

In this Appendix, we report a few additional examples of spectra,
in Figs.~\ref{Fig:mass3}--\ref{Fig:mass5}. The choices of $\Delta$ are such as to include the entirety of
 the catalogue in Ref.~\cite{Elander:2022ebt}. We fix the indicative value  $g=5$ in all plots.
 Qualitatively, all these plots resemble at least one of those in the main body, though quantitative features may be amplified or suppressed by changes in $\Delta$.

\end{document}